\documentclass[12pt]{article}
\usepackage{color,amsmath,amssymb,exscale,psfrag,epsfig}
\usepackage{cite,color,url}
\usepackage[colorlinks=true
,urlcolor=blue
,anchorcolor=blue
,citecolor=blue
,filecolor=blue
,linkcolor=blue
,menucolor=blue
,linktocpage=true
,pdfproducer=medialab
]{hyperref}
\input epsf
\newcommand{\m}[1]{\marginpar{{\tiny *}} }
\newcommand{\Gslash}{{\not \!\! G}}

\def\bea{\begin{eqnarray}}
\def\eea{\end{eqnarray}}

\newcommand{\mtt}{$m_{t\bar t}$ }
\newcommand{\ttbar}{$t\bar t$ }
\newcommand{\qqbar}{quark-annihilation }
\newcommand{\fqq}{$f_{q\bar q}$ }
\catcode`\@=11
\def\lsim{\mathrel{\mathpalette\@versim<}}
\def\gsim{\mathrel{\mathpalette\@versim>}}
\def\@versim#1#2{\vcenter{\offinterlineskip
\ialign{$\m@th#1\hfil##\hfil$\crcr#2\crcr\sim\crcr } }}
\catcode`\@=12
\catcode`@=12
\begin{document}
\thispagestyle{empty}
\begin{flushright}
\end{flushright}
\vspace{0.3in}
\begin{center}
{\Large \bf Top-antitop resonance searches beyond 1 TeV} \\
\vspace{1.0in}
{\bf Ezequiel Alvarez$^a$, Juan Ignacio Sanchez Vietto$^a$ and Alejandro Szynkman$^b$}
\vspace{0.2in} \\
{\sl $^a$ CONICET, IFIBA and Departamento de F\'{\i}sica, FCEyN, Universidad de Buenos Aires, \\
Ciudad Universitaria, Pab.1, (1428) Buenos Aires, Argentina} \\
{\sl $^b$ IFLP, CONICET - Dpto. de F\'{\i}sica, Universidad Nacional de La Plata, \\ 
C.C. 67, 1900 La Plata, Argentina} \\
\end{center}
\vspace{1.0in}

\begin{abstract}
We perform a general parton level analysis for the search of heavy resonant states in the production of $t \bar t$ pairs at the LHC with an integrated luminosity of 30 fb$^{-1}$. We assume the existence of resonances that only couple to quarks and propose kinematic cuts in order to increase the amount of events produced through quark-annihilation. We study the interplay between different variables and their impact on the purity of the selected sample. We make focus on the longitudinal ($\beta$) and transverse ($p_T$) momentum of the $t \bar t$ pair, and the scattering angle ($\theta$) in the center of mass reference frame.  We observe that $\beta$ is replaced by $\theta$ as a suitable discriminating variable of quark-annihilation processes for invariant masses above 1 TeV. Finally, we illustrate the analysis with a gluon resonance of 1.5 TeV and show the improvement in the sensitivity of the signal when cuts on $\theta$ are imposed.
\end{abstract}

\vspace*{30mm}
\noindent {\footnotesize E-mail:
{\tt \href{mailto:sequi@df.uba.ar}{sequi@df.uba.ar},
\href{mailto:jisanchez@df.uba.ar}{jisanchez@df.uba.ar},
\href{mailto:szynkman@fisica.unlp.edu.ar}{szynkman@fisica.unlp.edu.ar}}}

\newpage

\section{Introduction}

With what seems to be the recent discovery of the Higgs boson \cite{higgs}, the high energy physics enters into the solitude of a path that no one knows where or how it may found New Physics (NP).  Although a light Higgs requires either, fine-tuning, or NP at the TeV scale, the experimental results do not show any hint of new phenomena insofar.  As a matter of fact, {\it natural} Supersymmetry, which is one of the most popular theories to avoid fine-tuning and to simultaneously explain many other phenomena, is being highly constrained by direct searches at the LHC \cite{susy}.  

From a general theoretical point of view, one may expect to have NP effects in the detailed study of the heavier particles properties.  In particular, the top quark is the heaviest known particle and its properties have not been explored in depth insofar.  In fact, last years results from Tevatron \cite{tev} seemed to point to NP in the top forward-backward asymmetry.  However, recent results from the top charge asymmetry at LHC \cite{ac,ac2}, have contradicted most of the NP proposals except for very few models \cite{few} that have survived.  

Regardless of these recent results in top physics, in this article we consider the inclusive $pp \to t\bar t$ process where we study the $t\bar t$ invariant mass ($m_{t\bar t}$) spectrum as a sensitive observable to NP resonant phenomena.  This is due to a potential enhancement in the coupling of the top quark to the NP. In this work, we address the question of how to increase the sensitivity of this observable in the case of NP that couples to quarks and not to gluons.  The reason for this study is that at the LHC most ($\gtrsim 75\ \%$ at $\sqrt{s}=8$ TeV and increasing with energy) of the $t\bar t$ events have a gluon in the initial state partons and, therefore, a possible NP as the stated above would be per-se diluted just because of the initial state partons of the event.  For the sake of brevity, from this point forward we call gluon-fusion any event with at least one gluon in the initial state. 

The goal of this article is to propose kinematic cuts which enrich the quark-annihilation fraction ($f_{q\bar q}$) of a selected $t\bar t$ sample in order to increase the sensitivity to the stated NP in the $m_{t\bar t}$-spectrum.  Moreover, in light of the upcoming experimental results, we focus our work to the $m_{t\bar t} \gtrsim 1$ TeV region of energy.  This range of energies has two special features which determine the results in this work: {\it i)} we propose to use as one of the variables to discriminate quark-annihilation events, the center of mass scattering angle, which is useful for this purpose at high $m_{t\bar t}$; and {\it ii)} the lack of statistics at high $m_{t\bar t}$ requires a special selection criteria that with small cuts makes considerable increases in the sensitivity.  This latter is in contrast to the low energy regime, where strong cuts that yield a very large increase in the sensitivity are sought.

The search for kinematic cuts that increase \fqq in a given $t\bar t$ sample has been studied in the last years \cite{acsearches,hewett,bai,as,gr,afbseq} mainly to increase the sensitivity in the top charge-asymmetry at the LHC and very few in the $m_{t\bar t}$-spectrum \cite{mttseq,mttagashe}. These articles deal mainly with two different features in the production mechanisms that allow to distinguish $q\bar q$ production in $pp \to t\bar t$.  The first one is that valence quarks inside the colliding protons tend to have larger fraction of the proton momentum than gluons and anti-quarks.   Therefore, top pairs produced through quark-annihilation tend to be boosted along the pipe line.  This characteristic can be measured through the kinematic variable
\bea
\beta = \frac{\lvert p^z_t + p^z_{\bar t} \rvert}{E_t + E_{\bar t}},
\label{beta}
\eea
which ranges from $0$ for not boosted pairs to $1$ for maximum boosted pairs.  Notice for future purposes that the origin of this feature is in the proton parton distribution function (PDF).  The second feature which allows to isolate top pairs produced through quark-annihilation is that initial state gluons tend to produce more initial state radiation (ISR) than quarks.  This is an effect due to a larger numeric factor in the three gluon vertex in the QCD Lagrangian.  This characteristic can be measured through the transverse momentum of the top pair, 
\bea
p_T=| \vec p_T(t) + \vec p_T(\bar t) |.
\label{pt}
\eea
The larger is $p_T$, the more probable is to have a gluon in the initial state.  Notice that, in this case, the origin of this feature is pure QCD and is independent not only from the PDF, but also from the dynamics in the $t\bar t$ production process.  The usage of these two variables to increase the sensitivity to NP in the $m_{t\bar t}$-spectrum has been exploited in Refs.~\cite{mttseq,gr}.

There is a third variable which may serve to distinguish quark-annihilation processes. This is the center of mass scattering angle ($\theta$) of the top quark direction relative to the right moving parton.  This variable has been previously studied in Refs.~\cite{mttagashe,barger,maltoni}.  Since, as we show in the article, to take profit of this variable requires large $m_{t\bar t}$, we use it to study the upcoming experimental results where the $m_{t\bar t}$ region above $1$ TeV will be better analyzed.  This variable is related to the dynamics of the $pp\to t\bar t$ process and the spin of the particles.  The key feature in using the angular distribution is that for large $m_{t\bar t}$ the quark-annihilation processes have a smooth angular distribution, whereas gluon-fusion processes accumulate most of the events in the forward $|\cos (\theta)|\approx 1$ region. Or, to be more precise, the $t$-channel $gg \to t \bar t$ amplitude is the one that peaks the production in the forward region for relativistic tops.  Therefore, the angular distribution may also be understood as a discriminator of $s$- from $t$-channel contributions, as in dijet resonance searches \cite{dijet}.  

The aim of this work is to analyze simultaneously cuts in all three variables ($\beta,\ p_T$ and $\theta$) to enhance the sensitivity to resonant NP in the $m_{t\bar t}$-spectrum by increasing $f_{q\bar q}$.  Notice that, although all three variables have different origins, they have some degree of indirect correlation through the PDF's.  In any case --due to their different origin-- none of them can be expressed as a function of the other two.  The price to pay for increasing \fqq is to reduce the fraction of the original sample ($f_s$) that is selected for the analysis.  This reduction, which always yields an increase in the statistical uncertainty, may not worth the selection.  The critical point up to where it is convenient to cut the sample depends on how \fqq and the total uncertainty behave as a function of $f_s$.  Henceforth, it is crucial at this point also to take into account systematic uncertainties in the analysis in order to know if the selection cut is useful or not.  However, since a full realistic simulation of the systematic uncertainty would require a detailed simulation of the detector, which is beyond the scope of this work, we use a simpler model which is enough to lead us to the desired interplay between \fqq, $f_s$ and statistic and systematic uncertainties in order to decide whether the selection is suitable or not.  

Under this scenario, for the setup of the problem and its solution, it will be enough to consider $pp\to t\bar t$ processes up to parton level including ISR.  Hadronization, detector simulation and reconstruction would be useful only if one could perform a fully realistic detector simulation including its systematic uncertainties.  This is left for the experimental groups in case they consider to follow the guidelines in this work.  The purpose of this article is to show that exists an interplay between a selection cut on the three variables $\beta$, $p_T$ and $\theta$ and the statistic and systematic uncertainties, that yields an optimal cut which enhances the sensitivity of the $m_{t\bar t}$-spectrum to NP that couples to quarks.  We show along the article that this interplay, and therefore also the optimal cut, is strongly dependent on the energy of the process and the accumulated luminosity.  For instance, we show that for the 2012 LHC data, as the energy increases beyond $\sim 1$ TeV, strong cuts in $\beta$ should be gradually replaced by mild cuts in $\theta$. 

This article is divided as follows.  In next section we present the analytic results for the angular distributions in the Standard Model (SM) for quark-annihilation and $gg$-fusion.  We also study angular distribution of NP models which couple to quarks and not to gluons.  In section 3, we study the parton level variables $\beta$, $p_T$ and $\theta$ and their interplay with \fqq and $f_s$, as well as the relationships of these two with the uncertainties.  In section 4, we present an example of a resonant NP in $t\bar t$ production and show how the progressive cuts in the studied variables leads to the visibility of the resonant bump.  Section 5 contains a discussion on the results and previous works, and section 6 the conclusions. 

\section{Angular distribution in \ttbar production}

In this section we study analytically the leading order \ttbar production through the processes $q \bar q \to t \bar t$ and $gg \to t \bar t$ within the SM and for different possible resonant NP models. At high energy, events initiated by quarks have different angular distributions than those initiated by gluons. This is due to the $gg$-fusion amplitude where a top is exchanged in the $t$-channel producing \ttbar pairs mainly in the forward region. Therefore, one may impose kinematic cuts to the relevant angular variables in order to disentangle contributions from the two different production modes.

We begin analyzing the angular distribution of the two production mechanisms within SM. The expressions for the production squared amplitudes are given in the helicity basis in the initial parton center of mass frame. A sum over spin and color for the initial and final states as well as proper spin and color averaging factors which have not been included are assumed all along this section. For the \qqbar initial state, we have \footnote{As it is known, the study of the squared amplitudes is sufficient for the analysis developed in this section since the phase space in the two-body angular differential cross sections is independent of $\cos \theta$ itself.} \cite{parke}
\bea
\sum_{LL, RR} \lvert \mathcal M (q \bar q \to t \bar t) \rvert^2 &=& 8 g^4 \ (1-\beta_t^2) \sin^2 \theta , \label{qqLL} \\
\sum_{LR, RL} \lvert \mathcal M (q \bar q \to t \bar t) \rvert^2 &=& 8 g^4 \ (1+\cos^2 \theta). \label{qqLR}
\eea
\noindent
where $L$($R$) stands for left(right) helicity of the $t$ or $\bar t$, $\theta$ is defined as the angle between the direction of motion of the top and the initial parton moving to the right, and $\beta_t$ is the speed of the top quark. As it is apparent from these equations, the like-helicity ($LL$ and $RR$) $q \bar q \to t \bar t$ production is suppressed as $\beta_t \to 1$. This behavior is a consequence of the helicity conservation. 

For the $gg$ initial state we have \cite{parke}
\bea
\sum_{LL, RR} \lvert \mathcal M (gg \to t \bar t) \rvert^2 &=& 
\frac{16}{3} g^4 \ \frac{7+9 \beta_t^2 \cos^2 \theta}{(1- \beta_t^2 \cos^2 \theta)^2} \ (1-\beta_t^2) (1 + \beta_t^2 + \beta_t^2 \sin^4 \theta), \label{ggLL} \\
\sum_{LR, RL} \lvert \mathcal M (gg \to t \bar t) \rvert^2 &=& 
\frac{16}{3} g^4 \ \frac{7+9 \beta_t^2 \cos^2 \theta}{(1- \beta_t^2 \cos^2 \theta)^2} \ \beta_t^2 \sin^2 \theta (1 + \cos^2 \theta). \label{ggLR} 
\eea
For $\beta_t$ close to $1$, the common factor to both equations governs the growth of \ttbar production through $gg$-fusion as $\lvert \cos \theta \rvert \to 1$. Furthermore, this growth becomes stronger as $\beta_t \to 1$. Hence, we may expect that the angle $\theta$ becomes a more useful variable to distinguish \qqbar from $gg$ events for increasing values of the invariant mass of the \ttbar pair. Again, like-helicity contribution is suppressed in the limit $\beta_t \to 1$ as for \qqbar production, except for $\lvert \cos \theta \rvert \gtrsim 0.9$ where both production mechanisms contribute. This is understood because the $t$-channel that arises in this limit allows to flip the helicity of the tops.
Notice that at threshold ($\beta_t = 0$) both production mechanisms have an isotropic angular distribution. Then, no angular distribution based discrimination between the two production mechanisms is expected close to the threshold.

\begin{figure}[!htb]
\begin{minipage}[b]{0.06\textwidth}
~
\end{minipage}
\begin{minipage}[b]{0.30\textwidth}
\begin{center}
\small{unlike-helicity}
\end{center}
\end{minipage}
\begin{minipage}[b]{0.30\textwidth}
\begin{center}
\small{like-helicity}
\end{center}
\end{minipage}
\begin{minipage}[b]{0.30\textwidth}
\begin{center}
\small{total}
\end{center}
\end{minipage}
\newline
\begin{minipage}[b]{0.06\textwidth}
\vspace{1cm}
\begin{center}
\small{$q \bar q$} \\
\end{center}
\vspace{.8cm}
\end{minipage}
\begin{minipage}[b]{0.30\textwidth}                                
\begin{center}                                                            
\includegraphics[width=0.9\textwidth]{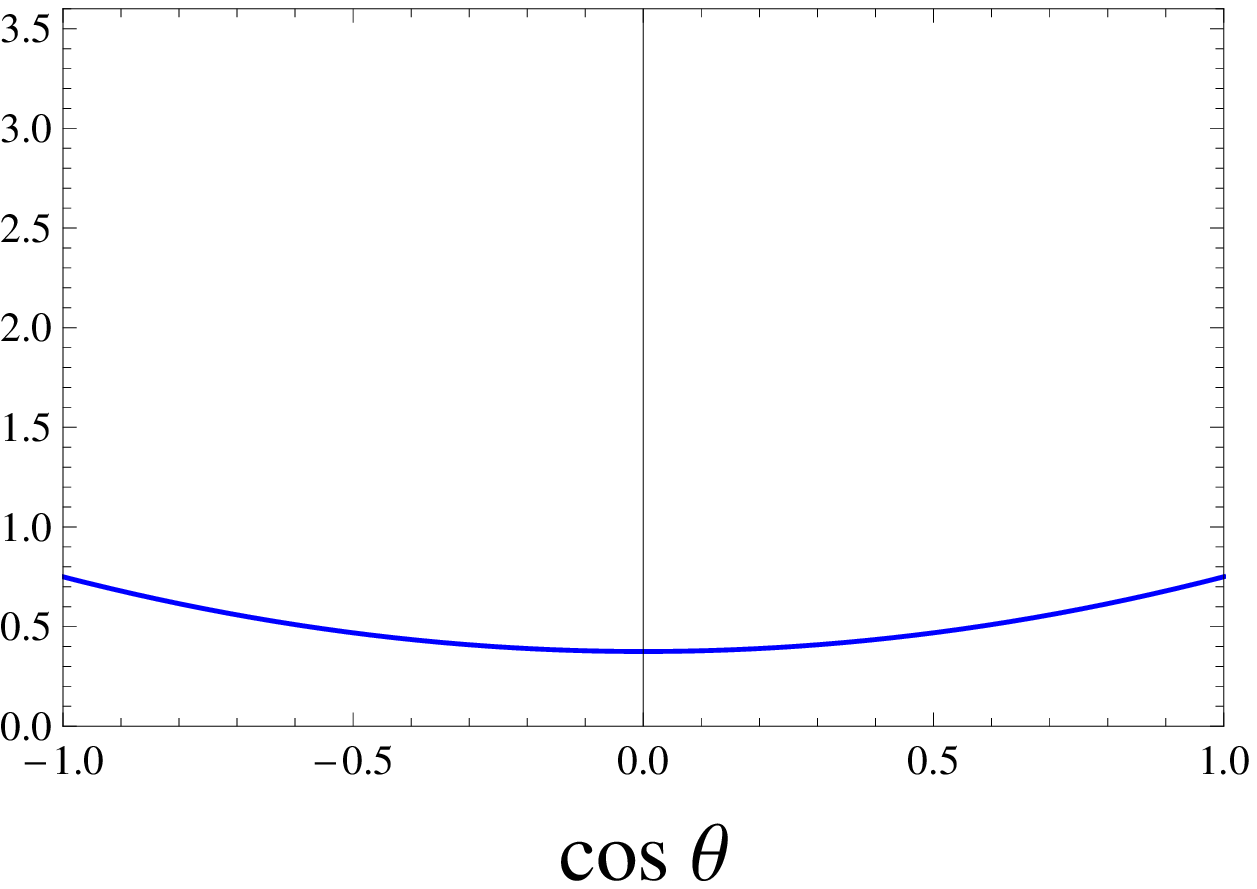}
\end{center}
\end{minipage}
\begin{minipage}[b]{0.30\textwidth}
\begin{center}
\includegraphics[width=0.9\textwidth]{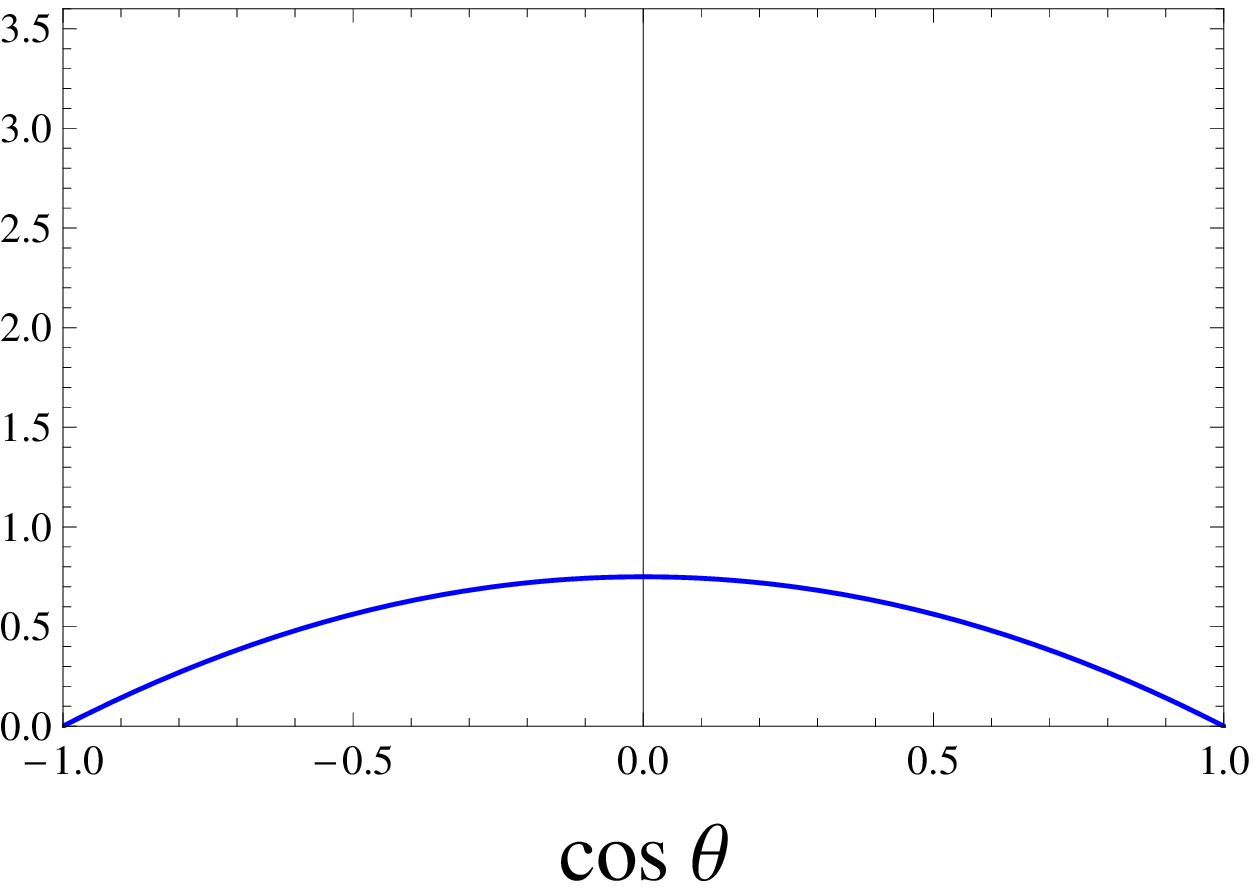}
\end{center}
\end{minipage}
\begin{minipage}[b]{0.30\textwidth}
\begin{center}
\includegraphics[width=0.9\textwidth]{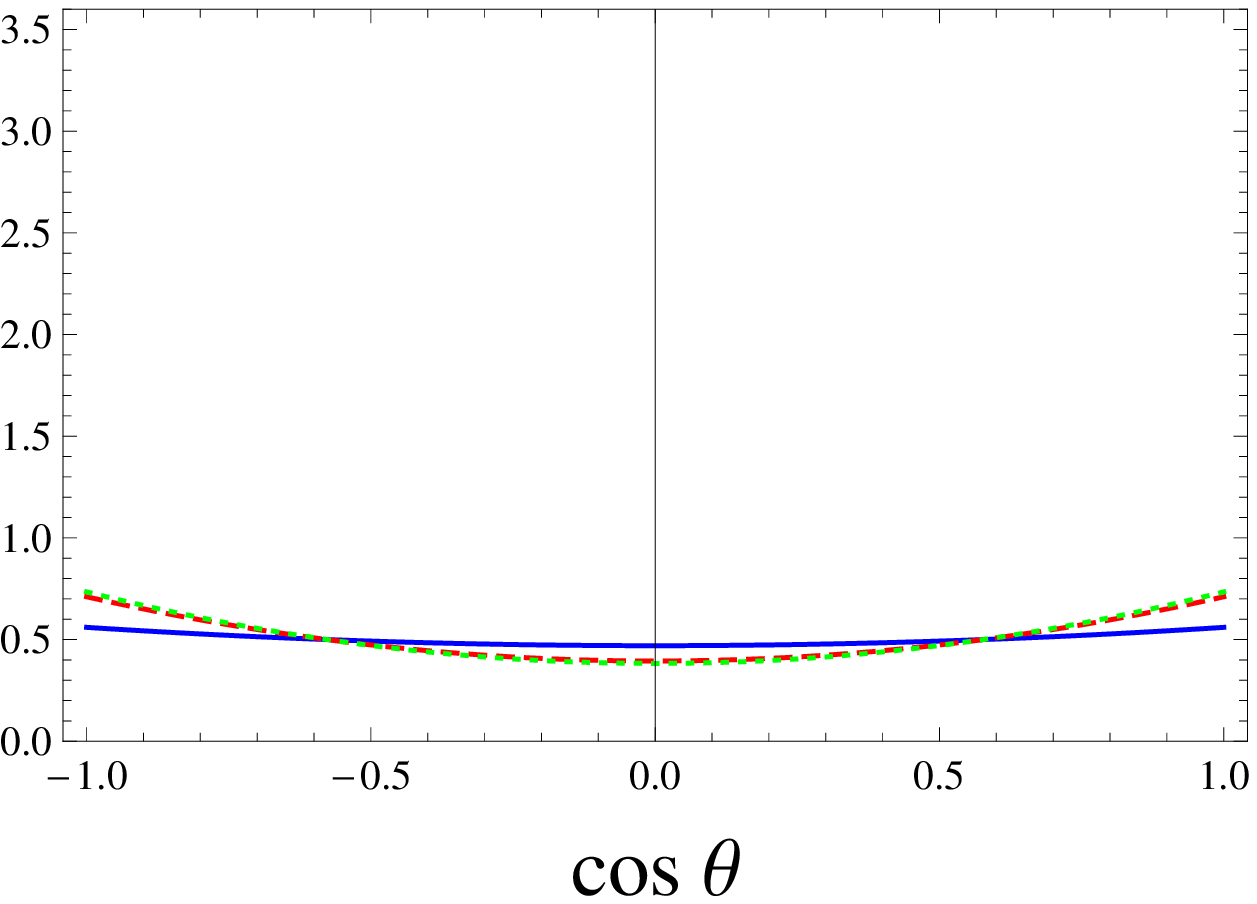}
\end{center}
\end{minipage}
\newline
\vskip .5cm
\begin{minipage}[b]{0.06\textwidth}
\vspace{1cm}
\begin{center}
\small{$gg$} \\
\end{center}
\vspace{.9cm}
\end{minipage}
\begin{minipage}[b]{0.30\textwidth}                                
\begin{center}                                                            
\includegraphics[width=0.9\textwidth]{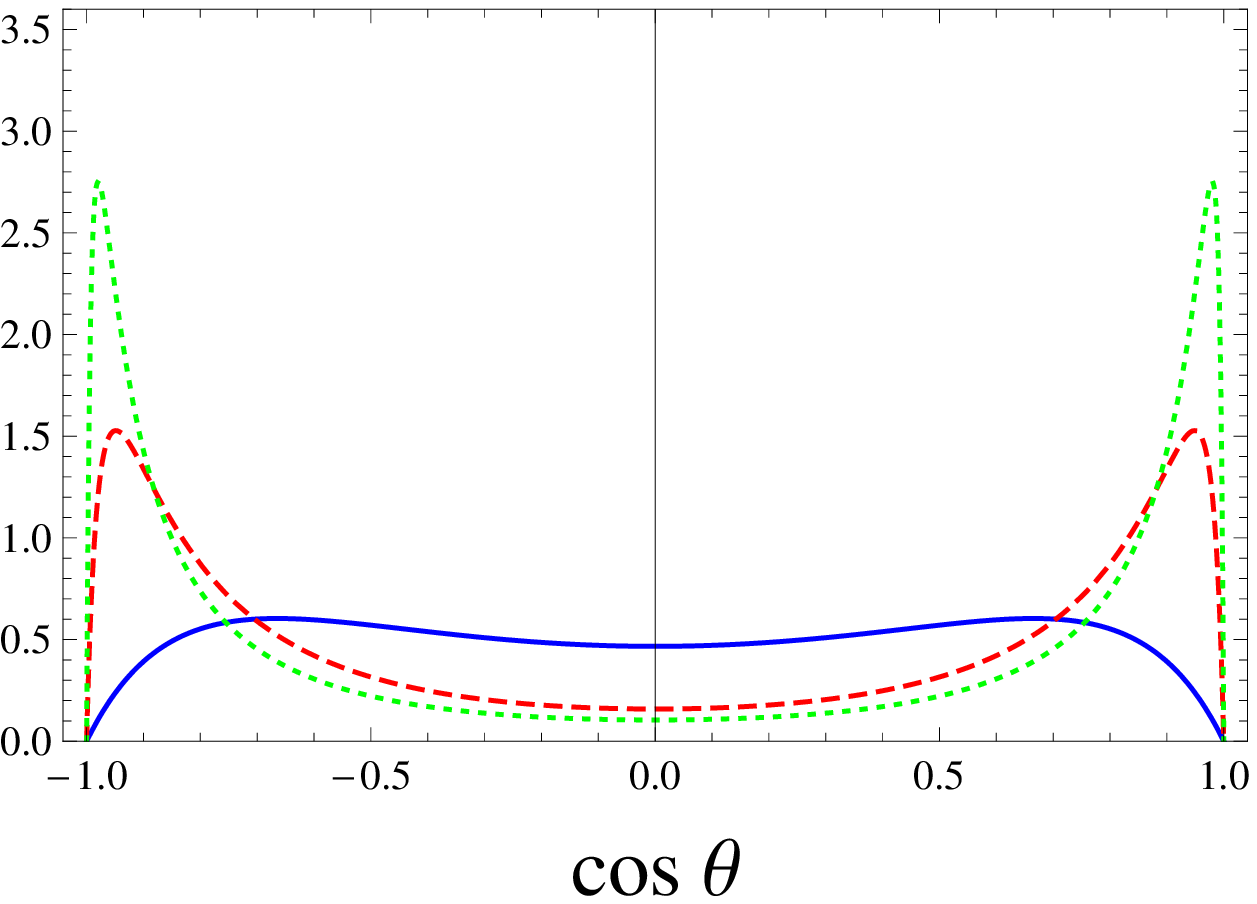}
\end{center}
\end{minipage}
\begin{minipage}[b]{0.30\textwidth}
\begin{center}
\includegraphics[width=0.9\textwidth]{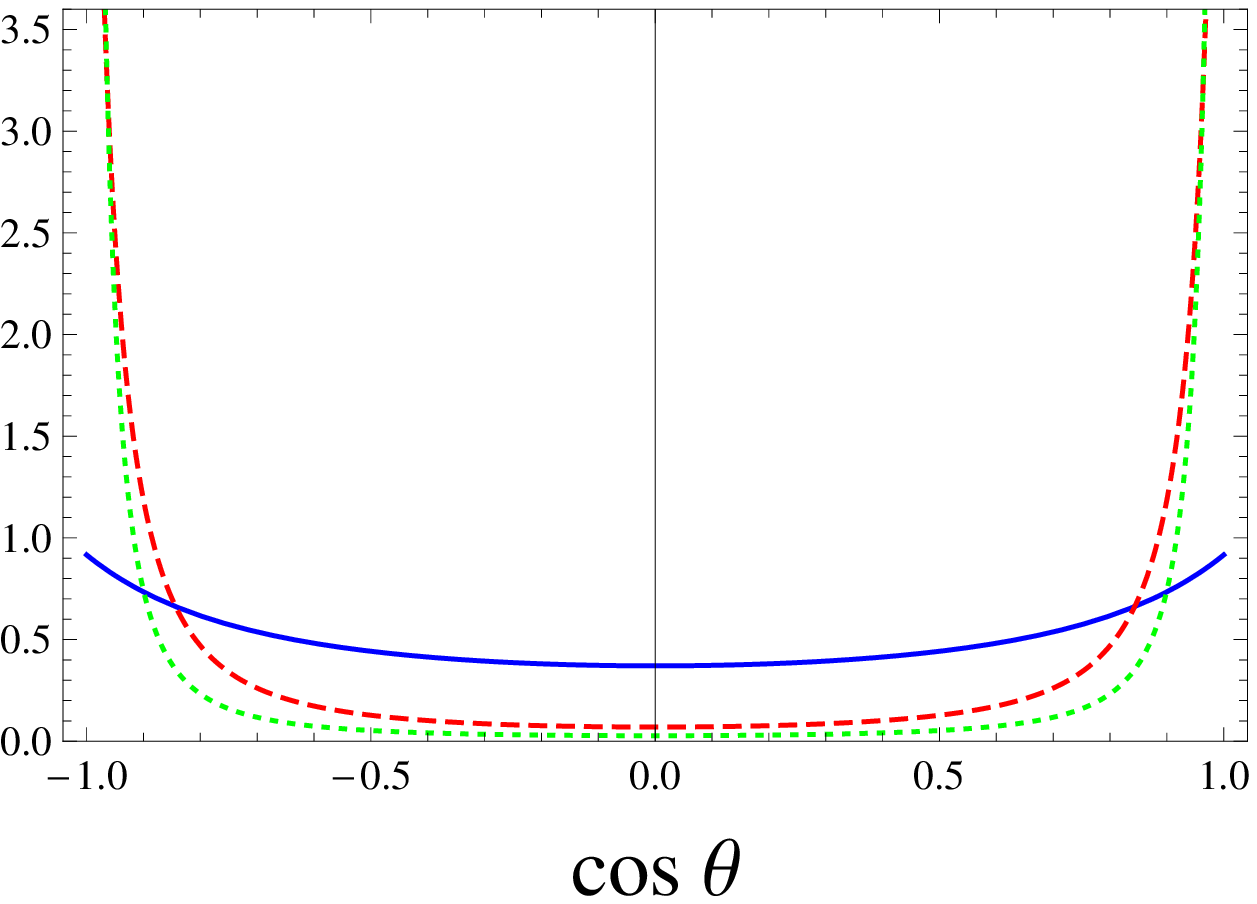}
\end{center}
\end{minipage}
\begin{minipage}[b]{0.30\textwidth}
\begin{center}
\includegraphics[width=0.9\textwidth]{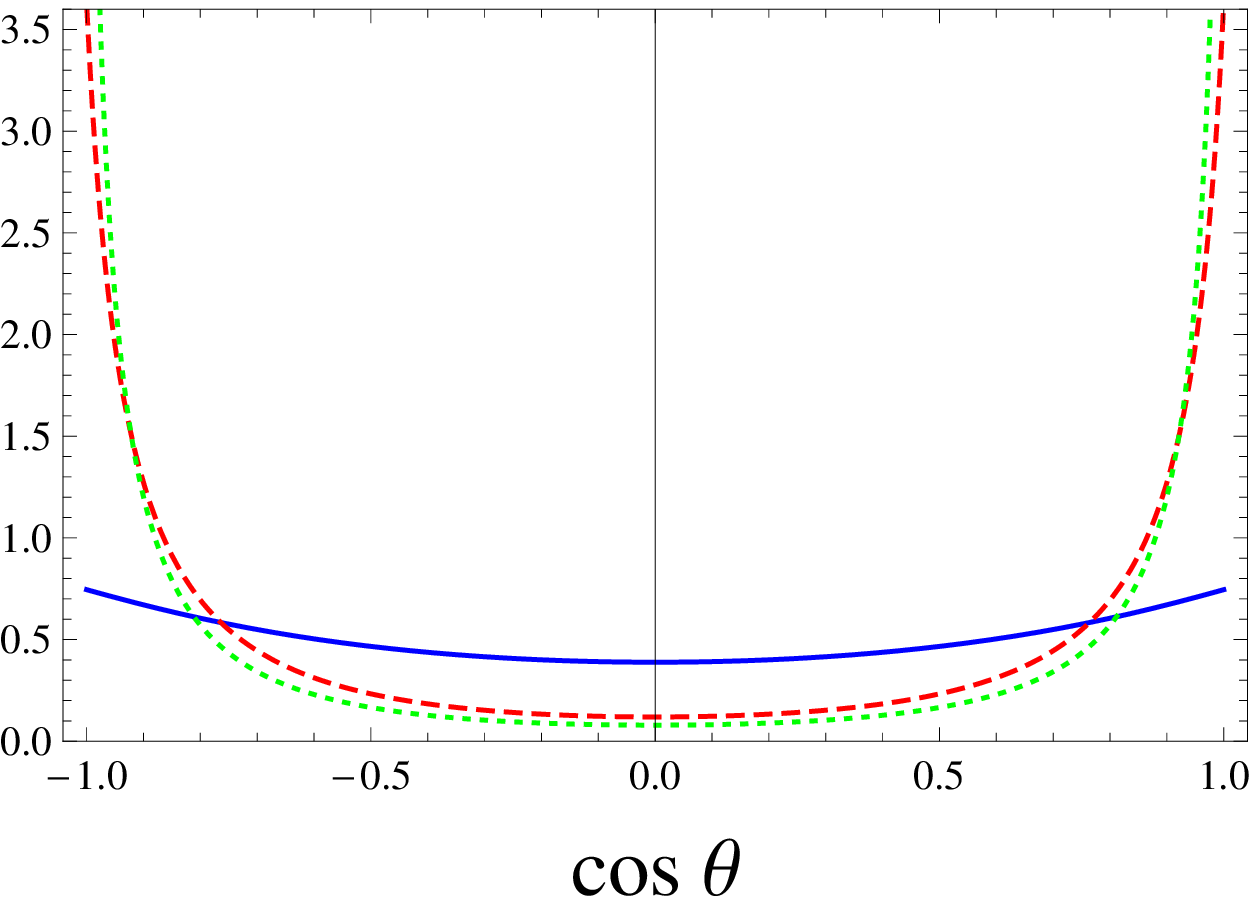}
\end{center}
\end{minipage}
\vspace{.5cm}
\newline
\caption{[color online] From left to right: normalized unlike-, like-helicity and total angular distributions for $q \bar q \to t \bar t$ (top) and $g g \to t \bar t$ (bottom) for three values of $\beta_t = 0.57$ (solid blue), $0.94$ (dashed red), $0.98$ (dot-dashed green). These correspond to $m_{t\bar t} = 425$ GeV, $1050$ GeV and $1700$ GeV, respectively. For \qqbar production the normalized like- and unlike-helicity angular distribution are independent of $\beta_t$ (notice that this only takes place when the angular distributions are normalized).}
\label{analytic}
\end{figure}

In Fig.~\ref{analytic} we show the normalized like- (left) and unlike- (center) helicity angular distributions from Eqs.~(\ref{qqLL})-(\ref{ggLR}) for three values of $\beta_t =0.57,0.94,0.98$ ($m_{t\bar t} = 425$ GeV, $1050$ GeV and $1700$ GeV, respectively). Also, we plot the total angular distributions for the sum of all final helicity states (right). Notice that for \qqbar production the normalized like- and unlike-helicity angular distribution are independent of $\beta_t$ since in Eq.(\ref{qqLL}) $\beta_t$ dependence enters just as a global factor and in Eq.(\ref{qqLR}) there is no dependence on $\beta_t$ at all. 
We can see that $q \bar q \to t \bar t$ production in an unlike-helicity state is more likely to be produced in the forward region whereas $q \bar q \to t \bar t$ production in a like-helicity state is mostly central. The suppression in the like-helicity $q \bar q \to t \bar t$ production for increasing values of \mtt results in a slight population of the forward region as we can see from the total $q \bar q \to t \bar t$ angular distribution. On the other hand, for $\beta_t \neq 0$  the unlike-helicity production for $gg \to t \bar t$ is turned on and both like- and unlike-helicity production begin to populate the forward region as $\beta_t$ increases. Hence, although both mechanisms tend to populate those regions for increasing values of $m_{t\bar t}$, $gg \to t \bar t$ production makes it further significantly due to the particular behavior of its total angular distribution as $\beta_t \to 1$. The comparison between $q \bar q \to t \bar t$ and $gg \to t \bar t$ productions in this limit is shown in the right-hand side column of Fig.~\ref{analytic}. These differences between the angular distribution of the two production mechanisms for large \mtt will prove to be useful to distinguish them. We may expect from these plots that as $\beta_t \to 1$ a loose upper cut in $|\cos\theta|$ results in a slight decrease of the total number of events  together with a sudden increment of the quark-annihilation fraction of the sample. This is in opposition to $\beta$
where one usually imposes tight cuts that end up removing a large amount of events from the sample in order to get an enhancement in the quark-annihilation fraction.

\begin{figure}[!htb]
\begin{minipage}[b]{0.32\textwidth}                                
\begin{center}          
\begin{center}
\small{spin-0}  \\
\end{center}                                             
\includegraphics[width=0.9\textwidth]{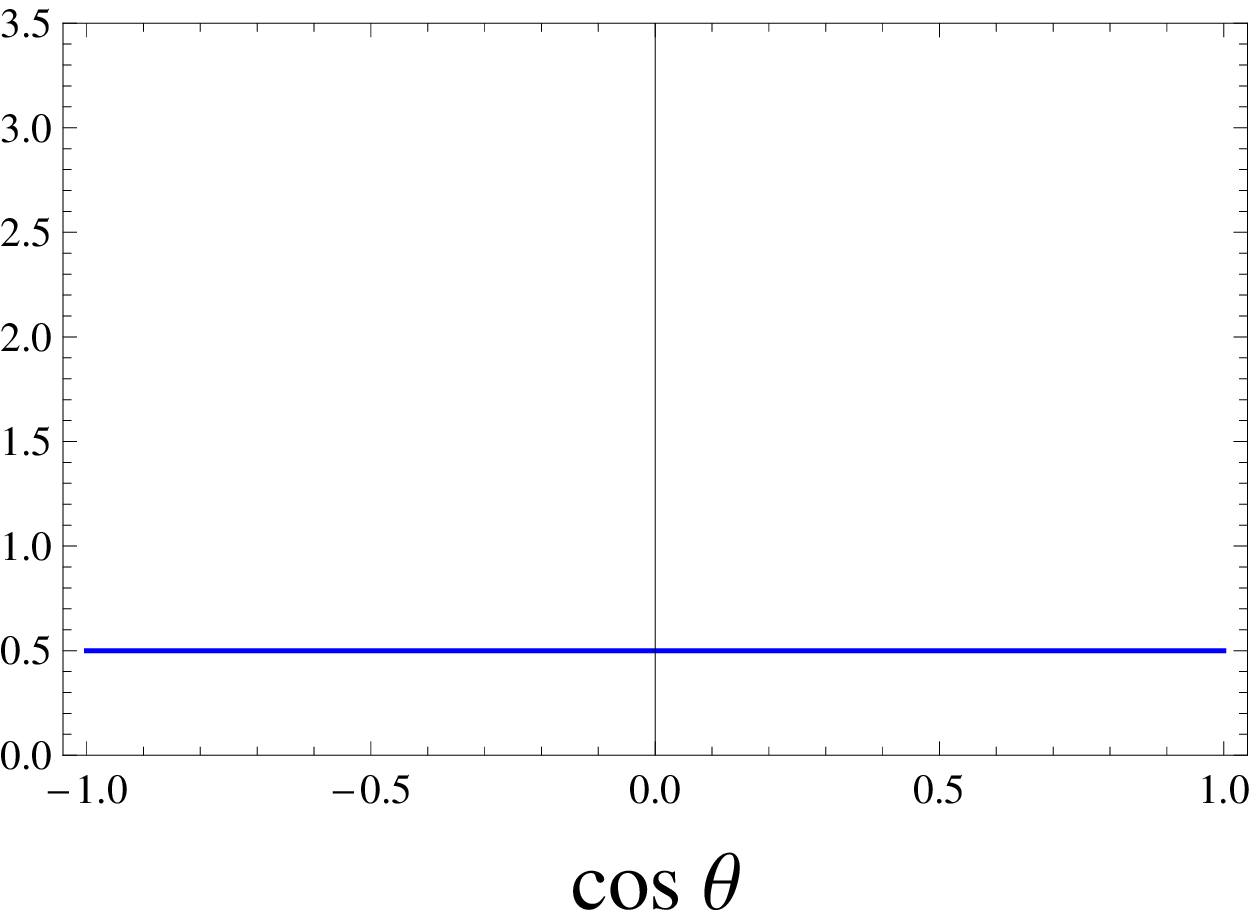}
\newline
(a)
\end{center}
\end{minipage}	
\begin{minipage}[b]{0.32\textwidth}                                
\begin{center}                                                           
\begin{center}
\small{spin-1}  \\
\end{center}                                             
\includegraphics[width=0.9\textwidth]{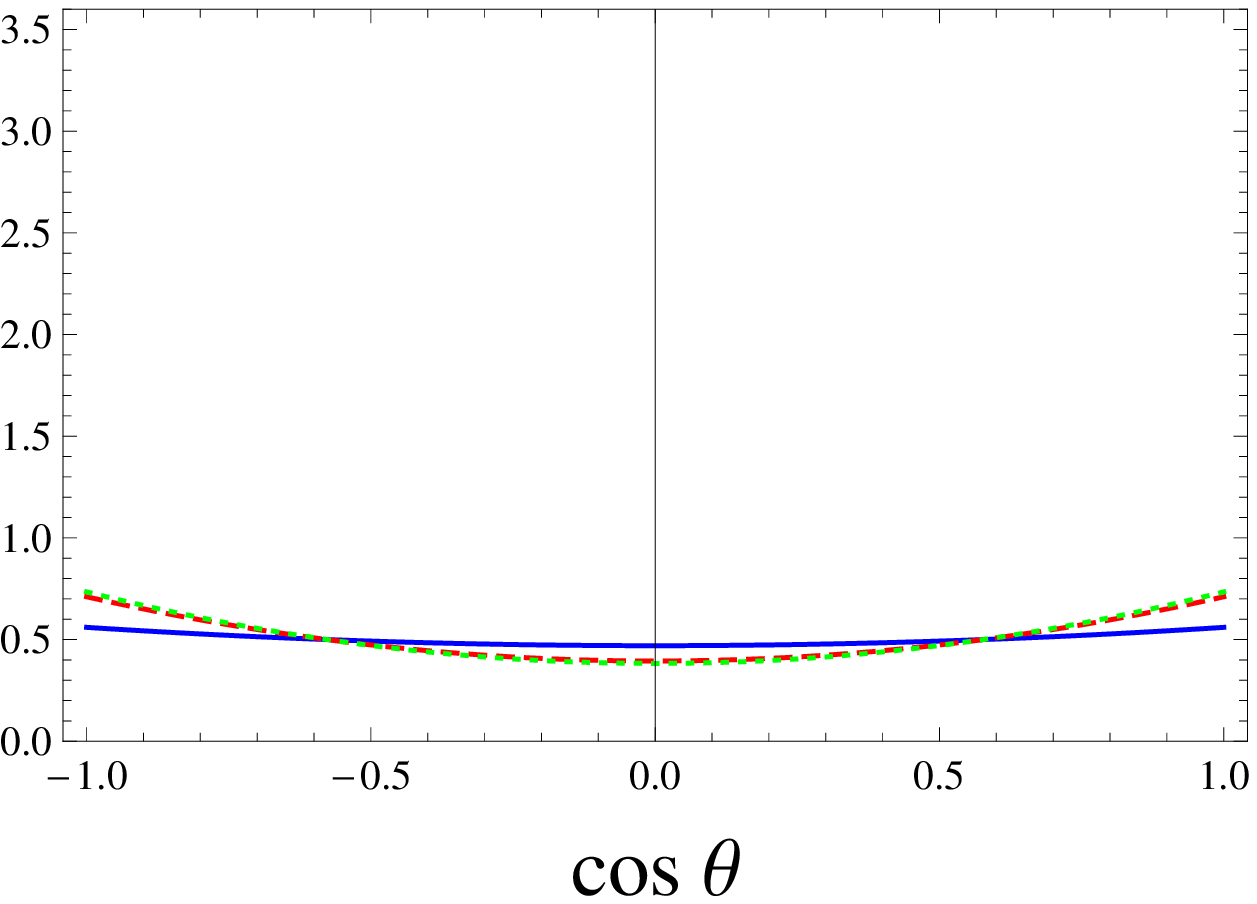}
\newline
(b)
\end{center}
\end{minipage}
\begin{minipage}[b]{0.32\textwidth}                                
\begin{center}
\begin{center}
\small{spin-2}  \\
\end{center}                                                                                                        
\includegraphics[width=0.9\textwidth]{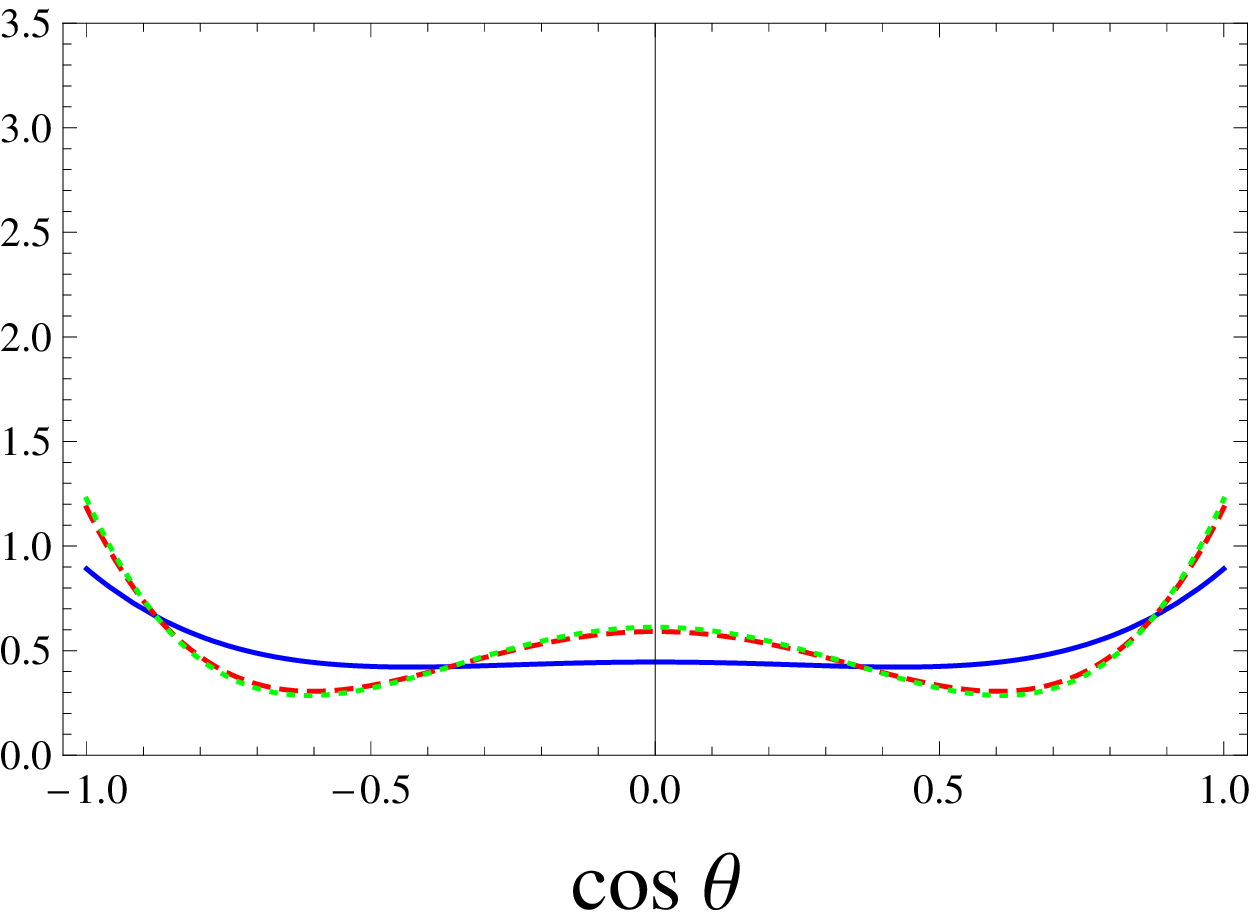}
\newline
(c)
\end{center}
\end{minipage}
\caption{\label{NP}[color online] Angular distribution of the $q\bar q \to t\bar t$ NP events through spin-0 (a), spin-1 vectorial or axial (b) or spin-2 (c) massive resonances. The different curves correspond to different $\beta_t$ as stated in Fig.~\ref{analytic}.} 
\end{figure}

In the presence of NP that couples only to quarks, the angular distribution of the $gg$-fusion events remains the same as in the SM. Therefore, the fraction of these events that are removed with a given cut on $\theta$ does not change with respect to the SM. For instance, we obtain from this simple parton level analysis that making a forward angular cut $\lvert \cos \theta \rvert < 0.85$ for $\beta = 0.98$, only $35\ \%$ of the $gg$-fusion initiated events pass the cut. This has to be compared with the fraction of SM quark-annihilation events that are left after the same selection, which is $\sim 80\ \%$.  Moreover, it is easy to verify that this selection cut is also useful for NP propagating in the $s$-channel.  We test this with three benchmark cases where the NP corresponds to spin-0, spin-1 or spin-2 massive resonances as is shown in Fig.~\ref{NP}. Since the spin-0 colorless resonance (Fig.~\ref{NP}a) does not interfere with the quark-annihilation QCD production and has no angular dependence, the same cut yields that $85\ \%$ of the NP quark-annihilation events are selected. For the spin-1 resonance (Fig.~\ref{NP}b) we choose a color octet $G'$ with vectorial couplings. The amplitude squared in this case is
\bea
\sum_{\text{spins}} \lvert \mathcal M(q \bar q \to g/G' \to t \bar t) \rvert^2 = A(s) (2-\beta_t^2 + \beta_t^2 \cos^2 \theta).  \label{M_NP}
\eea
The same angular dependence that the SM is obtained with $A(s)$ being a factor that depends on the resonance mass and width. This is consequence of the equal angular dependence of the QCD and NP amplitudes. Thus,  we also obtain that $80\ \%$ of this NP quark-annihilation events are left after the selection. Moreover, we find no difference with this and the case of an axial spin-1 color octet.  (Notice that in this case, if we had chosen $\theta$ as the angle between the top and the quark instead of a fixed beam direction, as is usual in forward-backward asymmetry studies, we would have obtained an asymmetric angular distribution for the $q \bar q \to G' \to t \bar t$ process, but the symmetric cut $\lvert \cos \theta \rvert < 0.85$ would have had the same effect.) Finally, in Fig.~\ref{NP}c we illustrate as an example a spin-2 colorless resonance \cite{spin2} which we restrict its couplings to quarks. In this case we find that $74\ \%$ of the NP events pass the selection. Thus, provided that the NP propagates in the $s$-channel, the cut in $\theta$ could be used to improve the quark-annihilation fraction and the sensitivity to resonant NP. In fact, although we find some slight differences in the efficiency of a given cut on $|\cos\theta|$ to keep quark-annihilation events from NP, we always obtain the same amelioration in reducing the dilution coming from $gg$-fusion events.  

\section{Using $\beta$, $p_T$ and $\theta$ to enhance the sensitivity to NP resonances}

We now discuss results from simulations of \ttbar events within the SM in order to find phase space regions where quark-annihilation is favored over gluon-fusion as the dominant production mechanism. We consider each kinematic variable separately to analyze them as potential filters of \qqbar processes. Finally, we perform random scans of cuts on these variables and show that to impose restrictions on $\theta$ is more efficient to isolate \qqbar production than to force limits on $\beta$ for the high \mtt region. 

We have simulated SM inclusive \ttbar production at the LHC@8TeV up to one extra jet with {\tt MadGraph5} \cite{mgme} and then showered it with {\tt Pythia} \cite{pythia}. To avoid double counting, we have matched the matrix element to the parton shower through the MLM scheme \cite{mlm} implemented in {\tt MadGraph5}. This generation procedure allows to include nontrivial $p_T$ distributions. It is beyond the scope of this work to deal with final states, therefore we have assumed a simplified selection cut $|\eta_{t, \bar t }| \leq 2.5$ (for future purposes, it is worth noticing that $\eta = 2.5$ corresponds to $\cos(\theta) = 0.987$, for the case where both laboratory and center of mass frames coincide). The prediction for the inclusive production cross section has been obtained with {\tt MCFM} LHC@8TeV at NLO as 225.2 pb \cite{mcfm} and an overall selection efficiency for semileptonic \ttbar detection of $6\ \%$ \cite{ac2} has been used in the analysis. This strategy has ended up with a total of 400.000 \ttbar events at the 2012's 30 fb$^{-1}$.

\begin{figure}[!htb]
\begin{minipage}[b]{0.32\textwidth}
\begin{center}
\small{$M=400-450$ GeV} \\
(94200 events)
\end{center}
\end{minipage}
\begin{minipage}[b]{0.32\textwidth}
\begin{center}
\small{$M=1000-1100$ GeV} \\
(2200 events)
\end{center}
\end{minipage}
\begin{minipage}[b]{0.32\textwidth}
\begin{center}
\small{$M=1600-1800$ GeV} \\
(120 events)
\end{center}
\end{minipage}
\newline
\begin{minipage}[b]{1\textwidth}                                
\begin{center}                                                            
\includegraphics[width=1\textwidth]{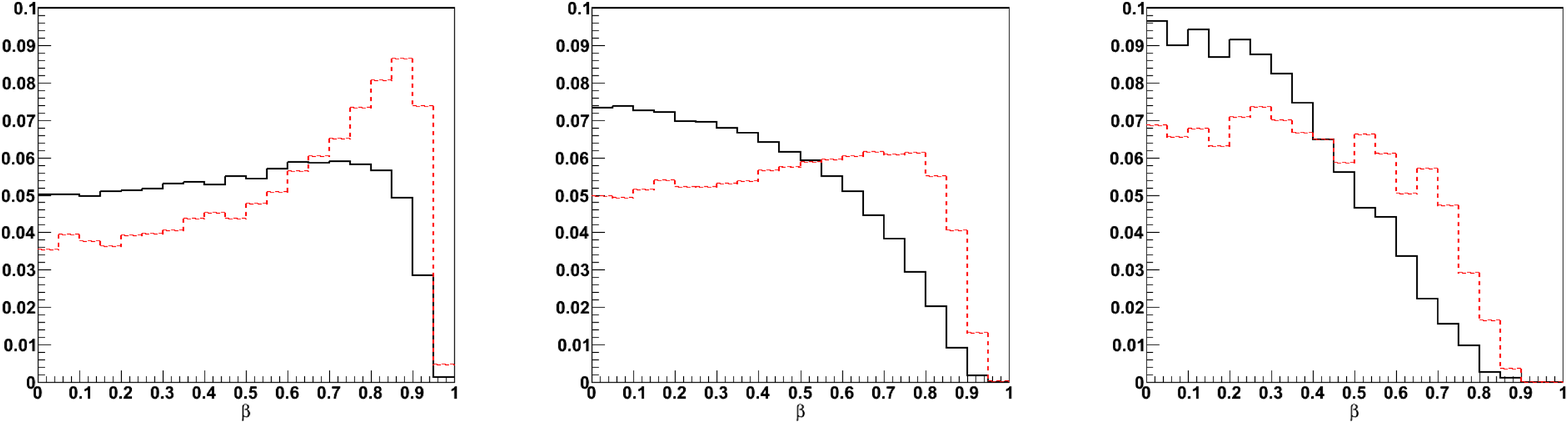}
\end{center}
\end{minipage}
\begin{minipage}[b]{1\textwidth}
\begin{center}
\includegraphics[width=1\textwidth]{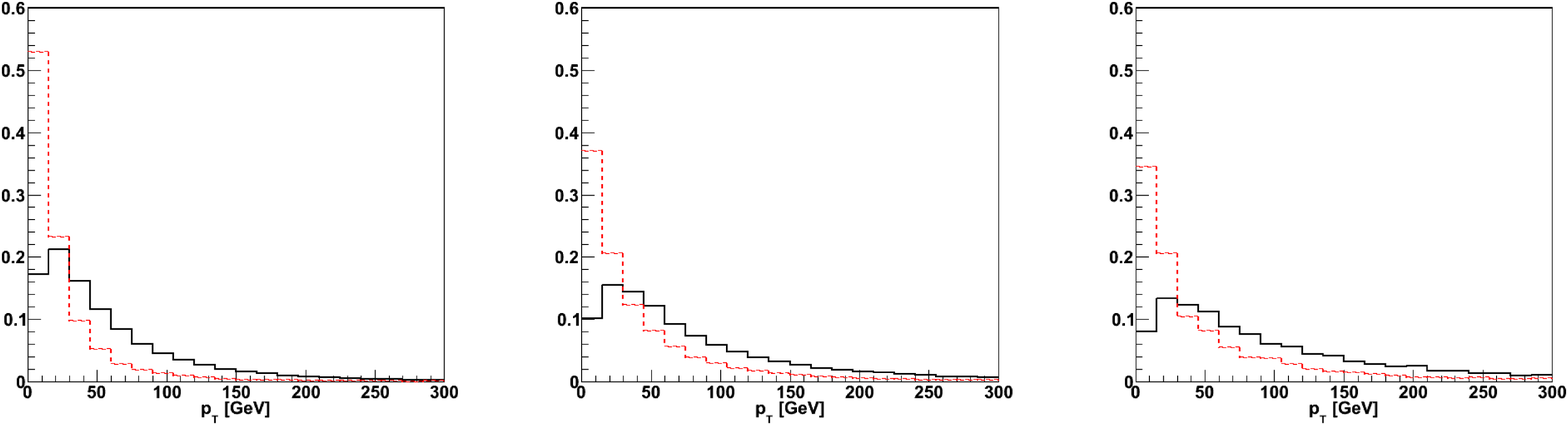}
\end{center}
\end{minipage}
\begin{minipage}[b]{1\textwidth}
\begin{center}
\includegraphics[width=1\textwidth]{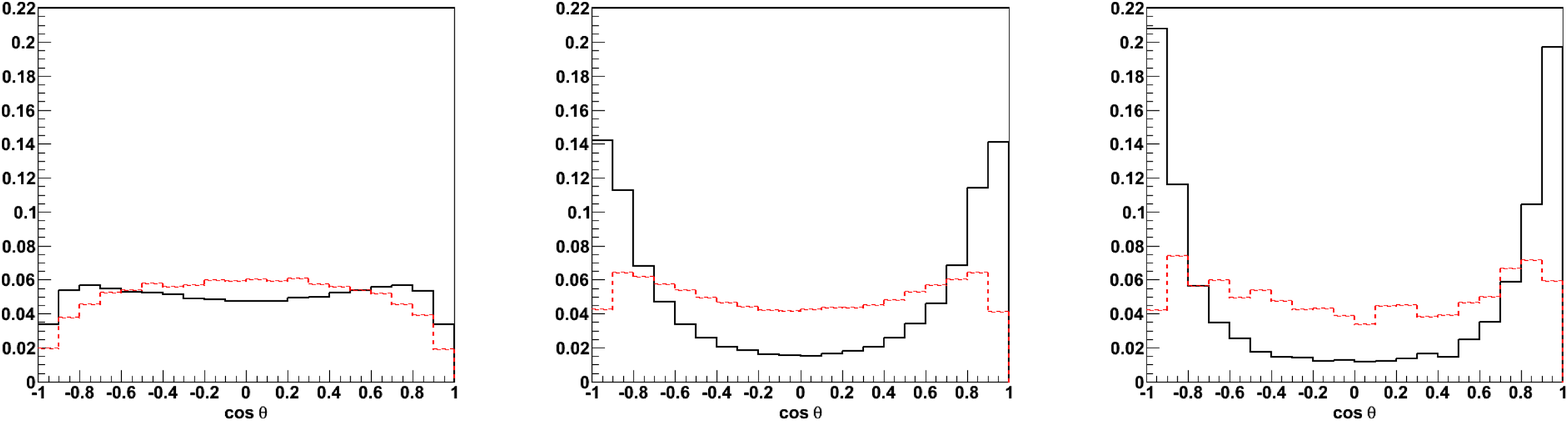}
\end{center}
\end{minipage}
\caption{[color online] Differential \ttbar distribution for \qqbar (red dashed) and gluon-fusion (black solid) events for each one of the kinematic variables $\beta$, $p_T$ and $\theta$ in three invariant mass bins (400-450 GeV, 1000-1100 GeV and 1600-1800 GeV). All the distributions have been normalized to one.}
\label{histograms1}
\end{figure}

We show in Fig.~\ref{histograms1} the differential distribution of \qqbar (red dashed) and gluon-fusion (black continuum) for each one of the kinematic variables $\beta$, $p_T$ and $\theta$ for three different bins of \mtt (400-450 GeV, 1000-1100 GeV and 1600-1800 GeV) that will be used from now on all along the analysis. Notice also that the central values of these bins correspond to the \mtt values selected in the previous section as referential cases. Since all the distributions have been normalized to one, the plots are not useful to compare the number of events initiated by \qqbar or gluon-fusion. Instead, they allow to evaluate whether a given cut is suitable for increasing the \qqbar fraction. Thus, we can see from the first row of the figure that \qqbar events tend to populate the high $\beta$ region more than gluon-fusion events. In particular, the peak in the distribution of \qqbar events at large $\beta$ in the 400-450 GeV can be understood according to the following. For large $m_{t \bar t}$, due to the suppression of the PDF's, it is easier the energy to be shared by the initial quark and anti-quark than having a configuration with large momentum difference between them. This situation is relaxed as \mtt decreases and it is more likely to find a quark with a large momentum. We conclude from the plots in the first row that a lower cut in $\beta$ always increases $f_{q\bar q}$.  By contrast, from the second row, we expect that an upper cut in $p_T$ may favor the \qqbar production. Finally, the angular distributions for both production mechanisms is depicted in the third row. As we have discussed in the previous section, both \qqbar and gluon-fusion events have essentially the same almost isotropic distribution in the 400-450 GeV bin. For increasing \mtt values, the angular distributions of both mechanisms starts to differentiate from each other. For the 1000-1100 GeV bin and even more remarkably for the 1600-1800 GeV bin, gluon-fusion events produce $t$ and $\bar t$ around the incoming direction of the initial partons whereas the events initiated by \qqbar do not have such a strong $\theta$ dependence. 

\begin{figure}[!htb] 
\begin{minipage}[b]{0.3\textwidth}
\begin{center}
\tiny{$M=400-450$ GeV}
\end{center}
\end{minipage}
~ 
\begin{minipage}[b]{0.3\textwidth}
\begin{center}
\tiny{$M=1000-1100$ GeV}
\end{center}
\end{minipage}
~
\begin{minipage}[b]{0.3\textwidth}
\begin{center}
\tiny{$M=1600-1800$ GeV}
\end{center}
\end{minipage}
\newline      
\begin{minipage}[b]{0.3\textwidth}
\begin{center}
\includegraphics[width=1\textwidth]{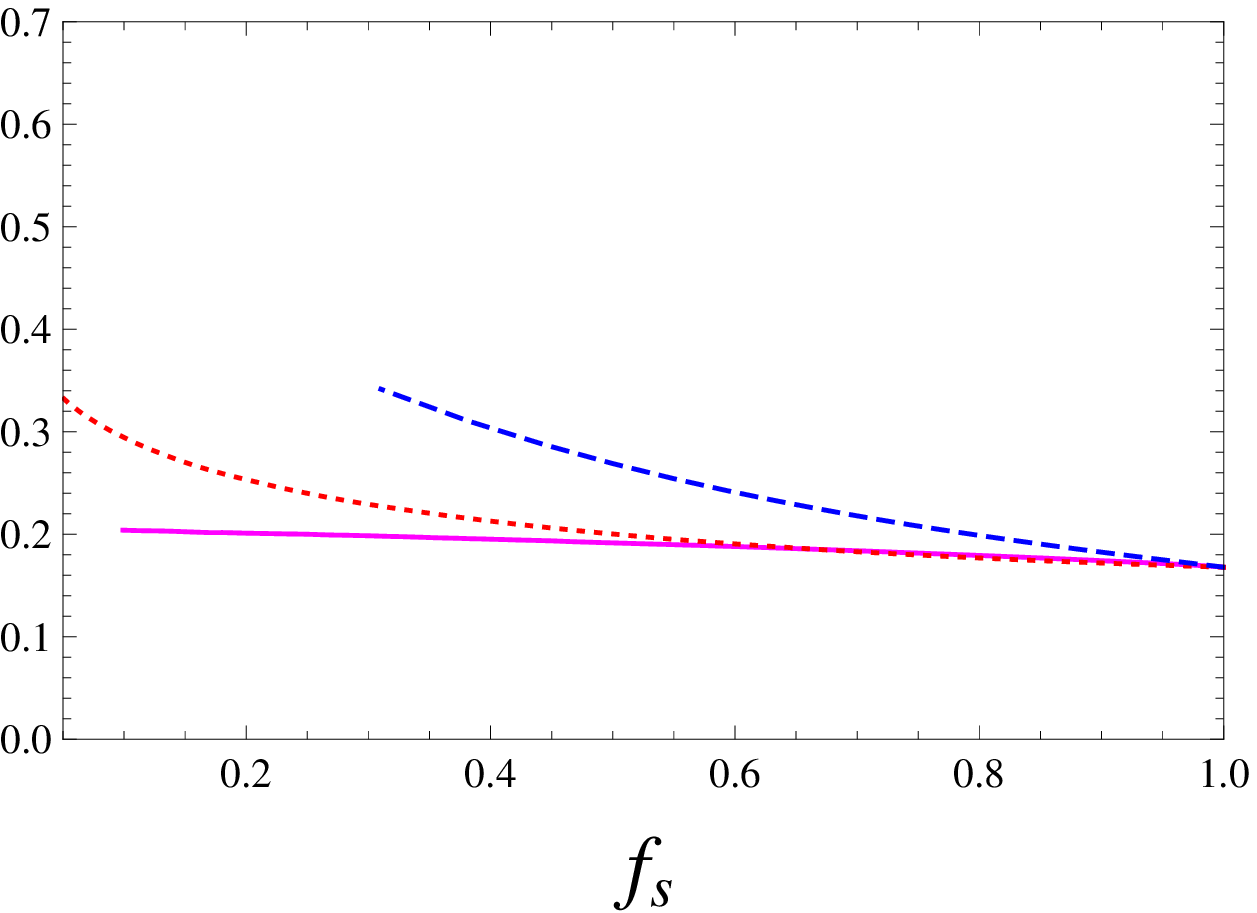}
\end{center}
\end{minipage}
~ 
\begin{minipage}[b]{0.3\textwidth}
\begin{center}
\includegraphics[width=1\textwidth]{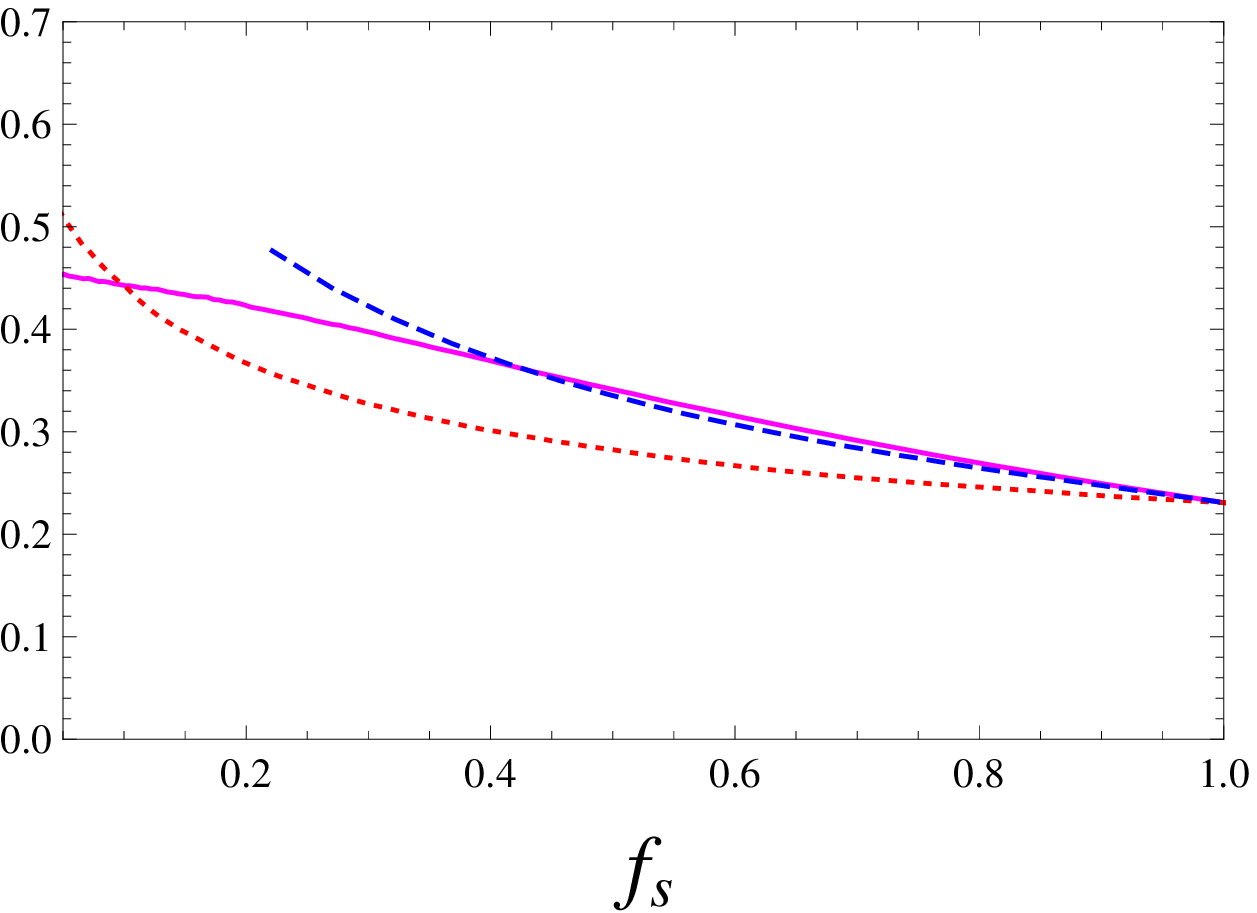}
\end{center}
\end{minipage}
~
\begin{minipage}[b]{0.3\textwidth}
\begin{center}
\includegraphics[width=1\textwidth]{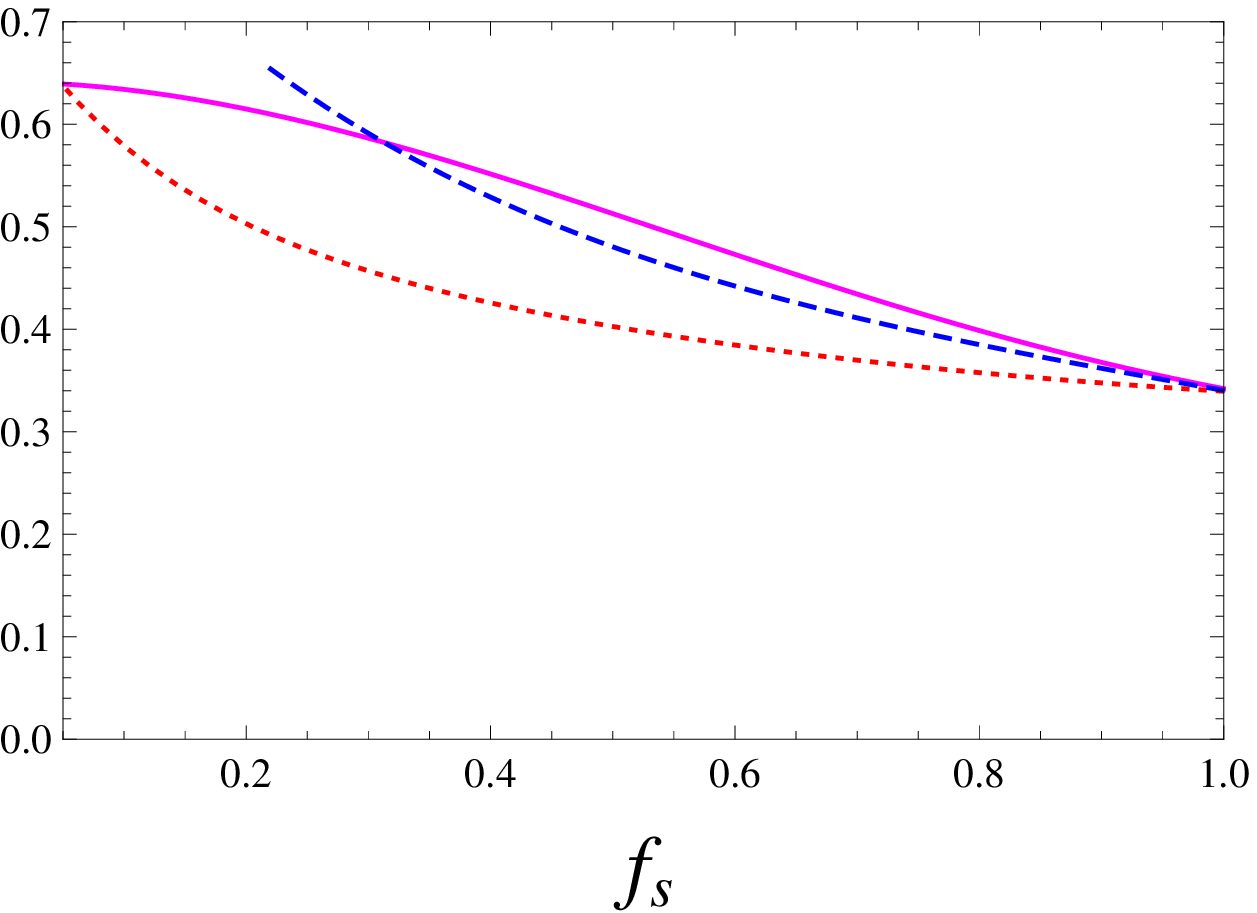}
\end{center}
\end{minipage}
\caption{[color online] \fqq (vertical axes) vs.~$f_s$ as the sample is cut in $\beta > \beta_c $ (red dotted), $p_T<p_{Tc}$ (blue dashed) or $|\cos\theta|<\cos\theta_c$ (magenta solid) for three invariant mass bins (400-450 GeV, 1000-1100 GeV and 1600-1800 GeV).  The range of the cuts are $\beta_c \in \left[0,0.95\right]$,  $p_{Tc} \in \left[20,300\right]$ GeV and $\cos \theta_c \in \left[0.1,1\right]$ }
\label{fq_vs_fs}
\end{figure}

We notice that the efficiency in performing cuts on $\beta$ and $p_T$ to disentangle \qqbar from gluon-fusion events is slightly sensitive to $m_{t\bar t}$. On the contrary, from the angular distributions we observe that gluon-fusion events populate the forward direction for increasing $m_{t \bar t}$ values and $\theta$ becomes a more convenient variable to discriminate \qqbar events. In order to go into greater detail on this matter, we have plotted in Fig.~\ref{fq_vs_fs} the \qqbar fraction (\fqq) vs. the fraction of events ($f_s$) that remain after a given cut  for each variable. For the 1600-1800 GeV bin, a larger \qqbar fraction with a less loss of events in the sample is achieved by setting cuts on $\theta$. Therefore, it is expected that cuts on $\theta$ to be more suitable to small samples of events as it is the case in resonance searches beyond $1$ TeV. In summary, cuts on $\theta$ may lead to larger values of $f_s$ in the relevant invariant mass region.  

So far we have considered each kinematic variable separately. Now, we combine cuts in the three variables with the aim to improve the \qqbar fraction. We perform then simultaneously a lower random cut $\beta > \beta_c$ and upper random cuts $p_T < p_{Tc}$ and  $|\cos \theta| < \cos \theta_c$. We have set $\beta_c \in \left[0,0.9\right]$,  $p_{Tc} \in \left[20,300\right]$ GeV and $\cos \theta_c \in \left[0.1,1\right]$. A cut on $\beta$ stronger than $0.9$ may suffer from large systematic uncertainties \cite{as} and $p_{Tc} = 20$ GeV \cite{ac2} corresponds to the minimal experimental sensitivity.

\begin{figure}[!htb] 
\begin{minipage}[b]{0.3\textwidth}
\begin{center}
\tiny{$M=400-450$ GeV}
\end{center}
\end{minipage}
~
\begin{minipage}[b]{0.3\textwidth}
\begin{center}
\tiny{$M=1000-1100$ GeV}
\end{center}
\end{minipage}
~
\begin{minipage}[b]{0.3\textwidth}
\begin{center}
\tiny{$M=1600-1800$ GeV}
\end{center}
\end{minipage}
\newline      
\begin{minipage}[b]{0.3\textwidth}
\begin{center}
\includegraphics[width=1\textwidth]{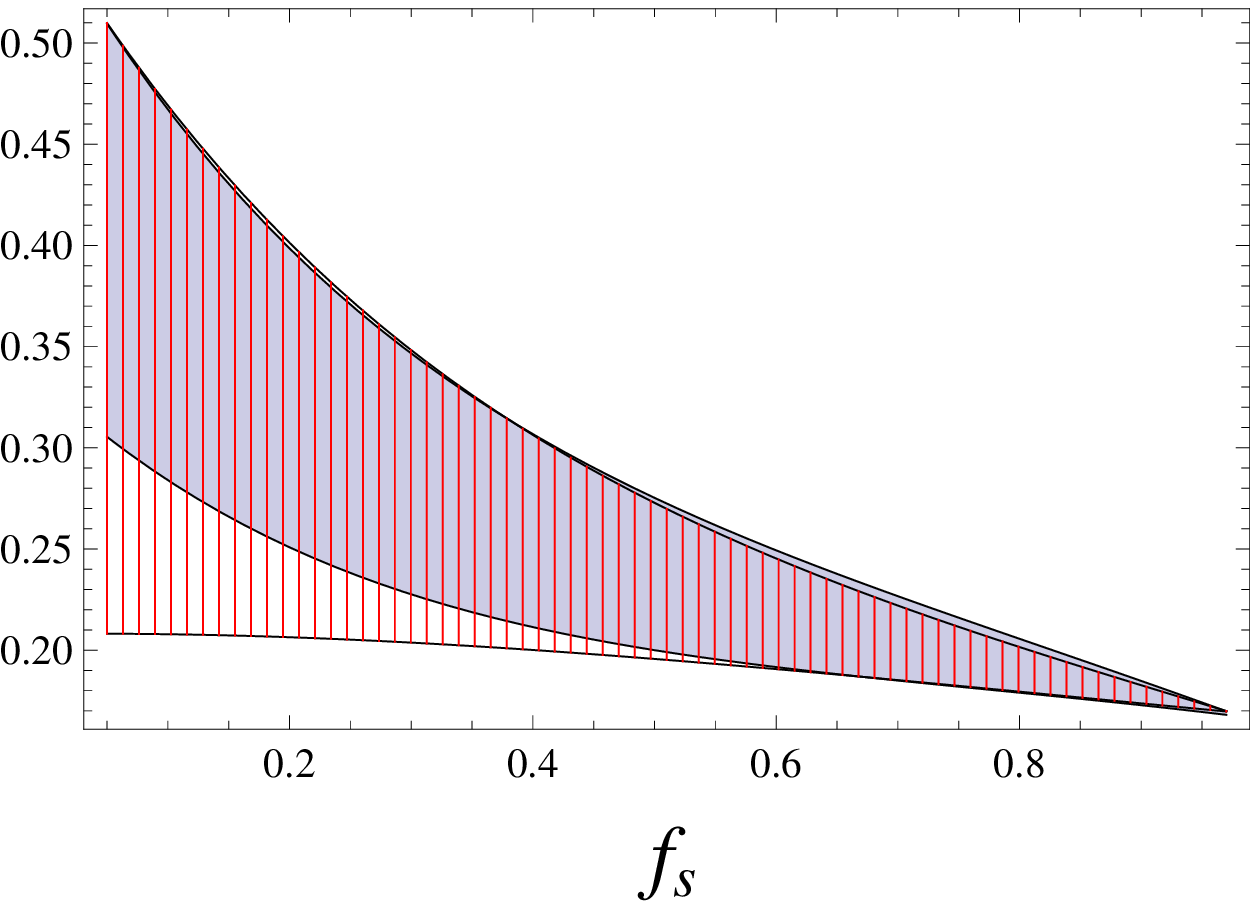}
\end{center}
\end{minipage}
~
\begin{minipage}[b]{0.3\textwidth}
\begin{center}
\includegraphics[width=1\textwidth]{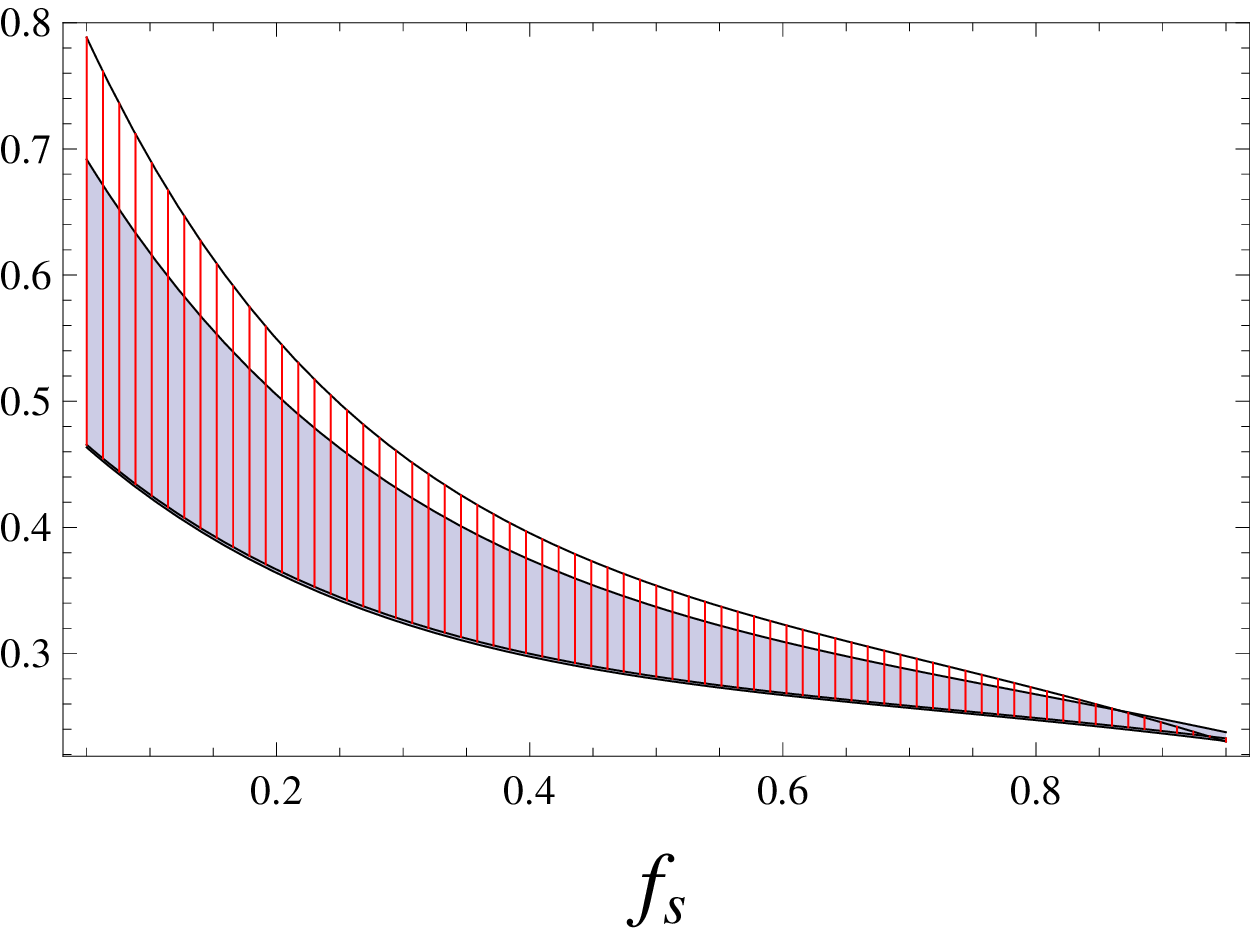}
\end{center}
\end{minipage}
~
\begin{minipage}[b]{0.3\textwidth}
\begin{center}
\includegraphics[width=1\textwidth]{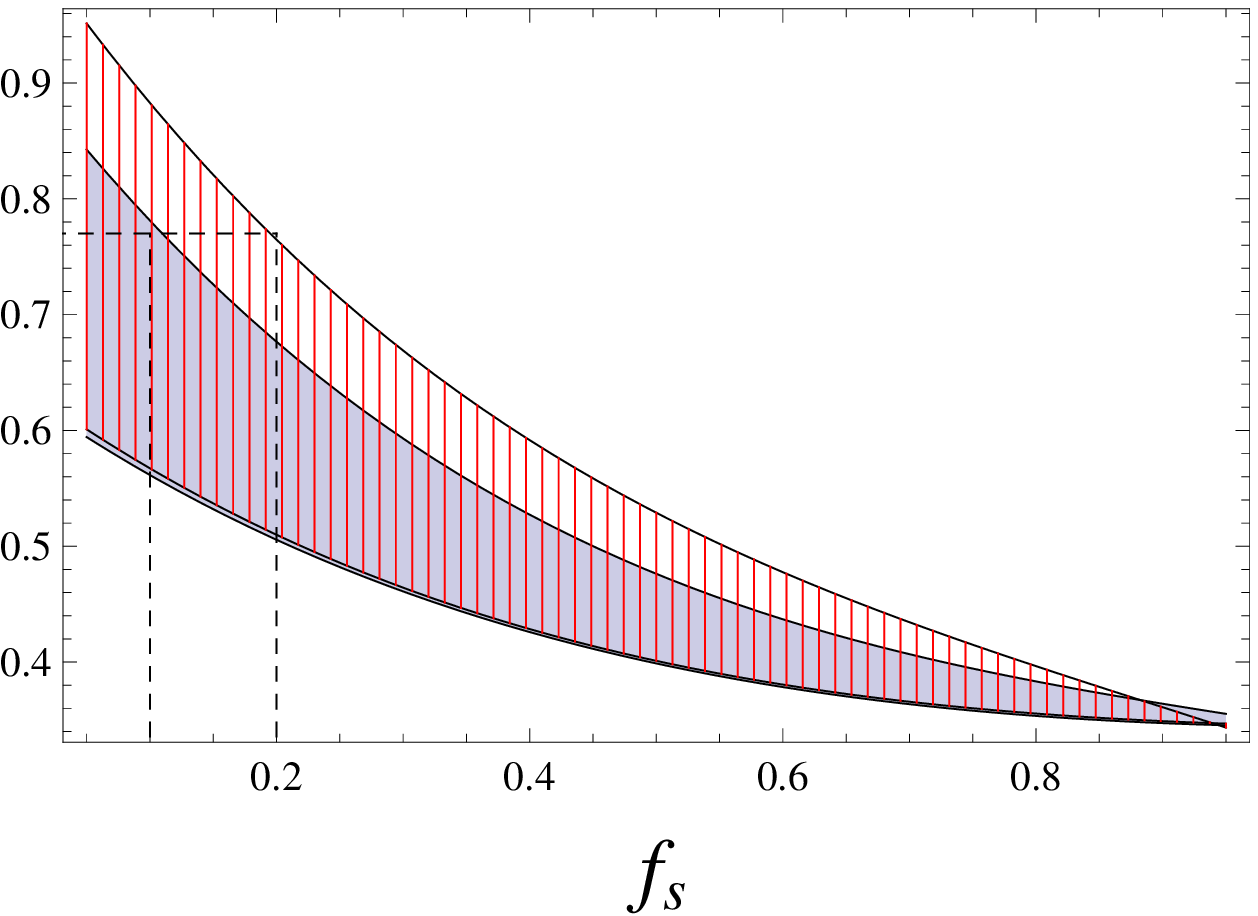}
\end{center}
\end{minipage}
\caption{[color online] \fqq (vertical axes) vs.~$f_s$ for a scan on the cuts on $\beta$, $p_T$ and $\theta$ (vertical lines area), and on $\beta$ and $p_T$ (shaded area) (no cut on $\theta$) for three invariant mass bins (400-450 GeV, 1000-1100 GeV and 1600-1800 GeV).}
\label{fq_vs_fs.scanning}
\end{figure}

In Fig.~\ref{fq_vs_fs.scanning} we show \fqq vs. $f_s$ for a scan on the three variables (vertical lines area), and only in $\beta$ and $p_T$ (shaded area) (namely, no cut on $\theta$ is demanded). We have plotted for the three invariant mass bins 400-450 GeV, 1000-1100 GeV and 1600-1800 GeV. As expected from the discussion given in the previous section, an enhancement in \fqq is obtained at large invariant masses when we combine cuts on all the three variables instead of only imposing cuts on $\beta$ and $p_T$. Moreover, we have also checked that, for the 1600-1800 GeV bin, the maximum \fqq for a fixed $f_s$ is reached without any cut on $\beta$. The independence on the cut on $\beta$ is consistent with the behavior observed in Fig.~\ref{fq_vs_fs} within the same \mtt range. Therefore, in order to end up with $f_s$ as large as possible for a given value of $f_{q\bar q}$, we conclude that performing cuts on $\theta$ is more efficient than restricting $\beta$. This is an important observation since $\beta$ is usually thought as a central kinematic variable in phenomenological analyses that aim at achieving \qqbar and gluon-fusion discrimination \cite{acsearches,hewett,bai,as,gr,afbseq}.

These SM results can be used to search for new physics in the \ttbar invariant mass spectrum. For the 1600-1800 GeV bin, simulated NP distributions follow the same behavior as in right plots in Fig.~\ref{histograms1}.  In particular, to impose upper cuts on $|\cos\theta|$ is still an efficient procedure to remove a considerable fraction of events initiated by gluon-fusion while keeping a large $f_s$, as discussed in previous section.  Therefore, the results of this section provide us with a guideline to look for new physics in the $m_{t\bar t}$-spectrum.

We consider now a possible matter of concern since, as the value of \fqq becomes larger, the number of events in the sample decreases with a logical increment in the statistical uncertainty. This may spoil the potential power of the whole analysis and, for resonance searches, we do not only need a large \fqq but also to have uncertainties under control. From Fig.~\ref{fq_vs_fs.scanning} we can see that, as $f_s$ gets reduced, there is always a set of cuts that improves $f_{q\bar q}$. If the systematic uncertainties dominate over the statistical ones, any cut enhancing \fqq always improves the sensitivity to new physics. However, if we are searching for resonances in the region beyond $1$ TeV, a reduction on the amount of events is expected and further cuts may lead immediately to a regime where both systematic and statistical uncertainties are competitive. Even if \fqq is large, the statistical uncertainties can spoil the sensitivity to NP \cite{mttseq}. Therefore, a regime where systematic and statistical uncertainties are of the same order has to be considered as a minimal necessary condition to be fulfilled.

It is beyond the scope of this work to perform an exhaustive analysis of the uncertainties and, for the sake of the discussion, we simply assume that the relative systematic uncertainty is constant. This is a good approximation as far as we do not push the cuts to extreme values. We model the systematic uncertainty in each bin as $\sigma^{syst} =  c N$, where $N$ is the number of events in the bin and $c$ is the relative systematic error in corresponding bin. A typical value for $c$ in a \ttbar-spectrum measurement without jet substructure top tagging is about $20\ \%$ ($c=0.2$) \cite{20p}.  (This number, and the selection efficiency, are expected to change if top tagging through jet substructure \cite{js} is used in the analysis.)  Thus, we get that the systematic and statistical uncertainties are of the same order provided that $N \sim 1/c^2 \approx \mathcal O(10)$. It is worth noticing at this point that if the systematic uncertainties are reduced ($c$ decreases) then the optimal number of events for a given bin increases in order to reduce the statistical uncertainties. Therefore, if the systematic uncertainties are reduced then the optimal selection cuts should be relaxed.   

We illustrate the above discussion with an example. At $8$ TeV with $30$ fb$^{-1}$, we expect in the SM 120 events in the bin of 1600-1800 GeV of invariant mass. As we have mentioned, the systematic and statistical uncertainties will be comparable for $N \approx 25$, i.e. $f_s \approx 0.2$. From Fig.~\ref{fq_vs_fs.scanning} we observe that for $f_s \approx 0.2$ there exists a set of cuts on $\beta$ (which turns out to be needless, $\beta > 0$), $p_T$ and $\theta$ that achieves $f_{q \bar q} \approx 0.75$. Whereas for cuts only on $\beta$ and $p_T$ we would need $f_s \approx 0.1$ to achieve the same $f_{q\bar q}$ (see the dashed line in Fig.~\ref{fq_vs_fs.scanning}c). Accordingly, we would obtain the same significance with a cut on $\beta$, $p_T$ and $\theta$ at 30 fb$^{-1}$ as with a cut only on $\beta$ and $p_T$ for 60 fb$^{-1}$. For this reference example, the set of cuts is $\beta \geq 0$, $p_T < 24$ GeV and $\lvert \cos \theta \rvert < 0.89$ for the former case, and $\beta > 0.38$ and $p_T < 20$ GeV for the latter. This case shows that to include $\theta$ as a discrimination variable is equivalent to double the luminosity.

 \section{Application of the studied cuts in a specific example}

In this section we apply the previously discussed cuts to a specific NP model in order to explicitly show the amelioration in the invariant mass spectrum sensitivity to a new resonance.  We study the effects of including cuts in different sets of $\beta$, $p_T$ and in $\theta$ to visualize the improvement in sensitivity as more variables are included in the cuts.  

\begin{figure}[!htb]
\begin{minipage}[b]{0.5\textwidth}                                
\begin{center}                                                           
\includegraphics[width=0.9\textwidth]{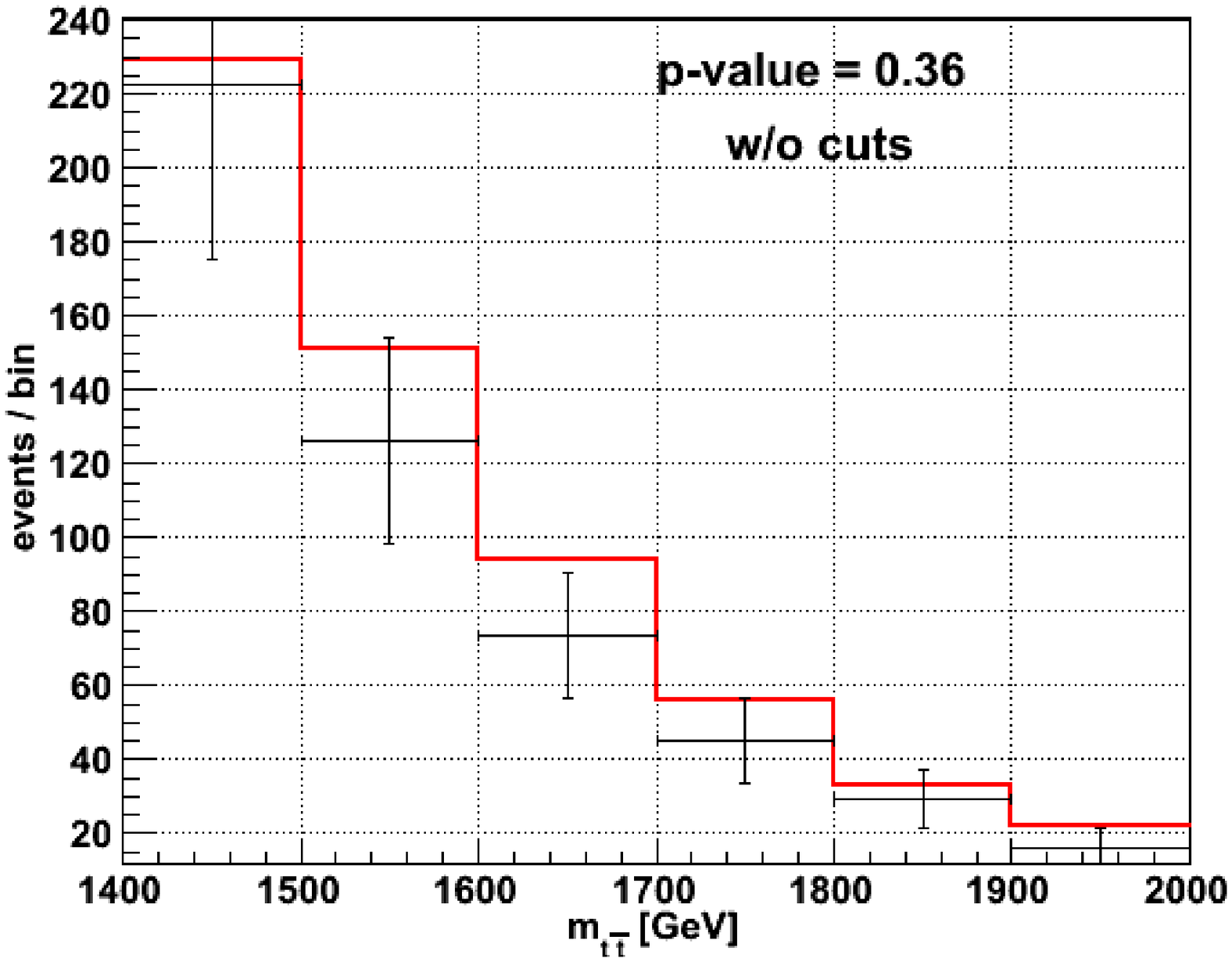}
\newline
(a)
\end{center}
\end{minipage}	
\begin{minipage}[b]{0.5\textwidth}
\begin{center}
\includegraphics[width=0.9\textwidth]{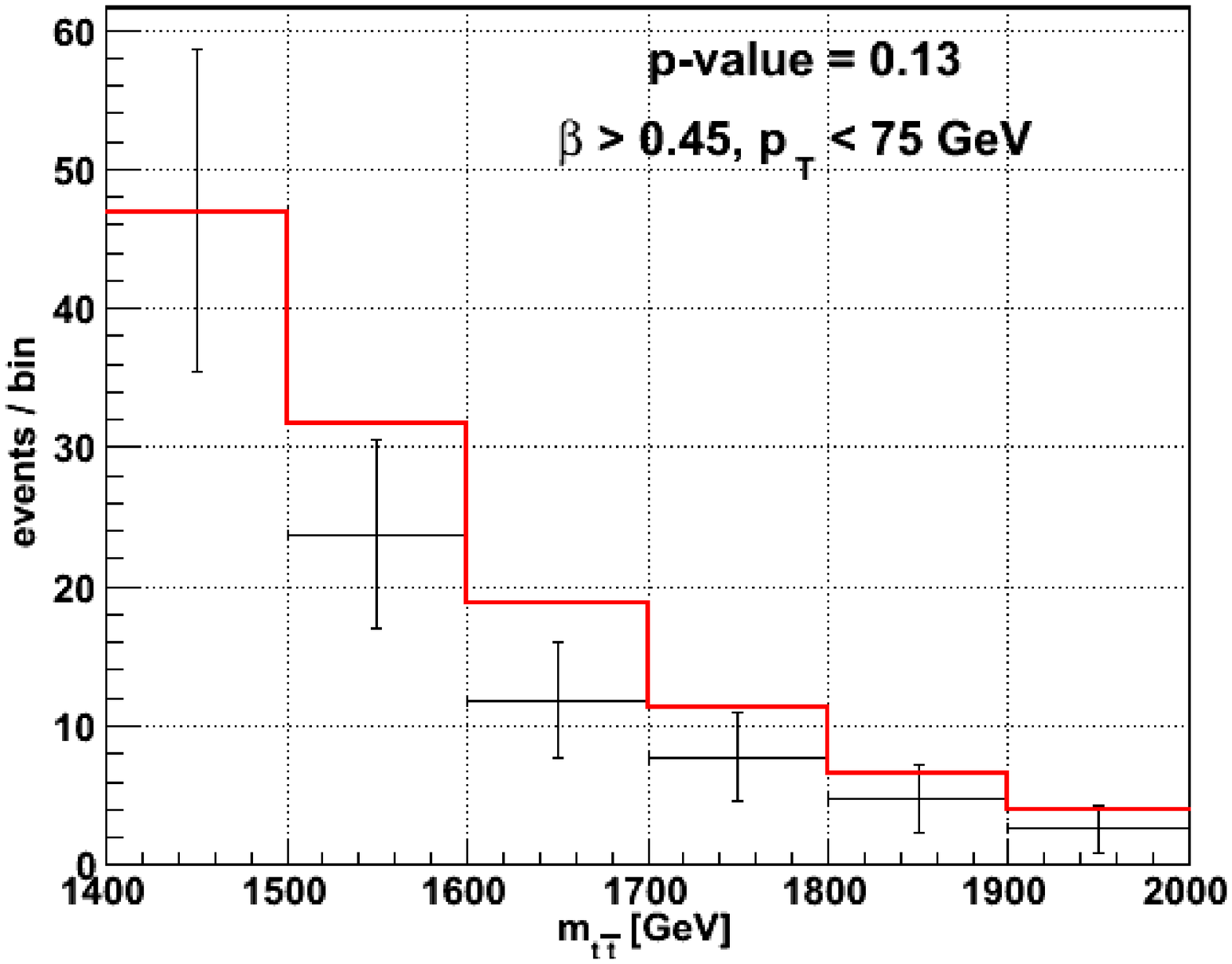}
\newline
(b)
\end{center}
\end{minipage}
\begin{minipage}[b]{0.5\textwidth}
\begin{center}
\includegraphics[width=0.9\textwidth]{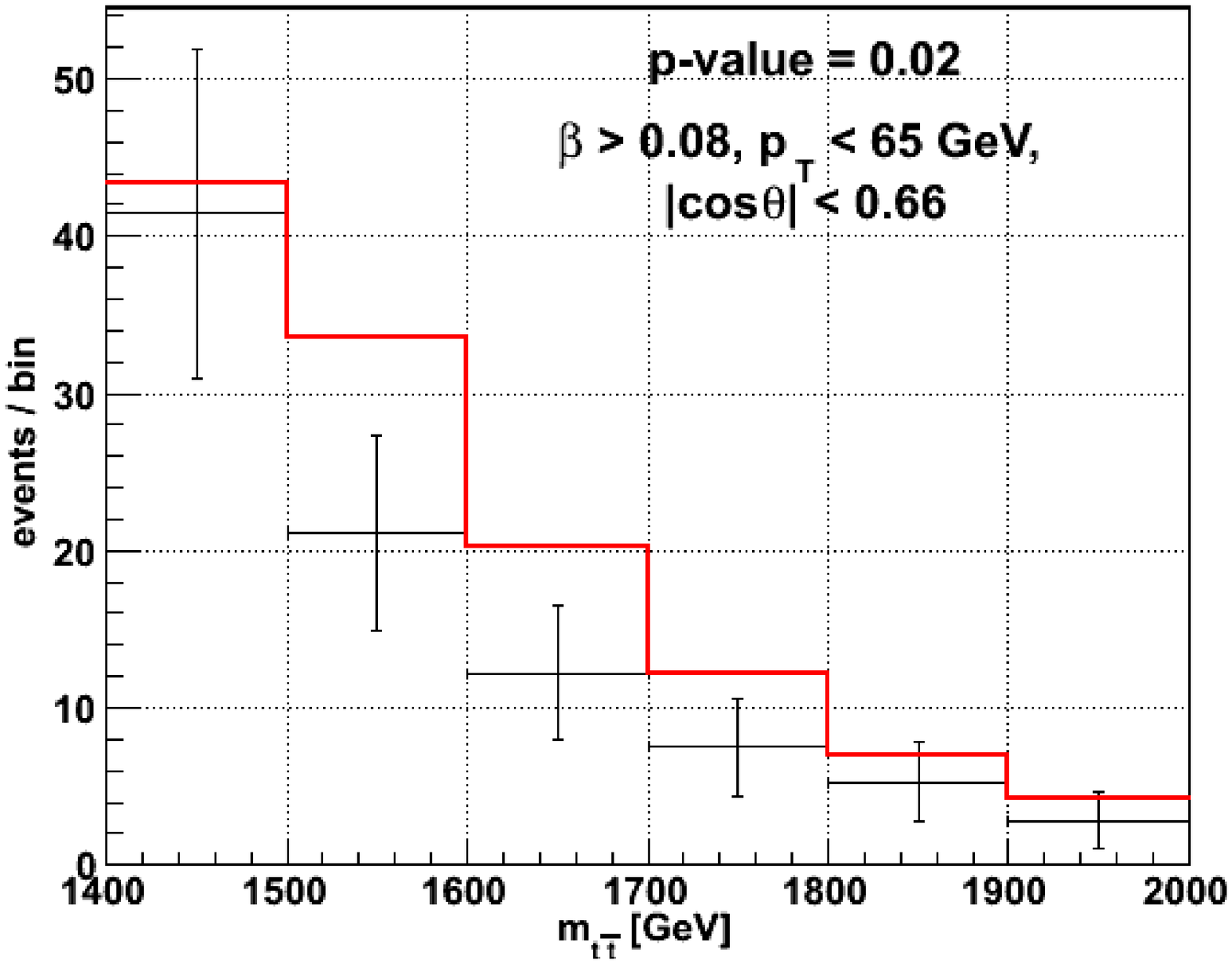}
\newline
(c)
\end{center}
\end{minipage}
\begin{minipage}[b]{0.5\textwidth}
\begin{center}
\includegraphics[width=0.9\textwidth]{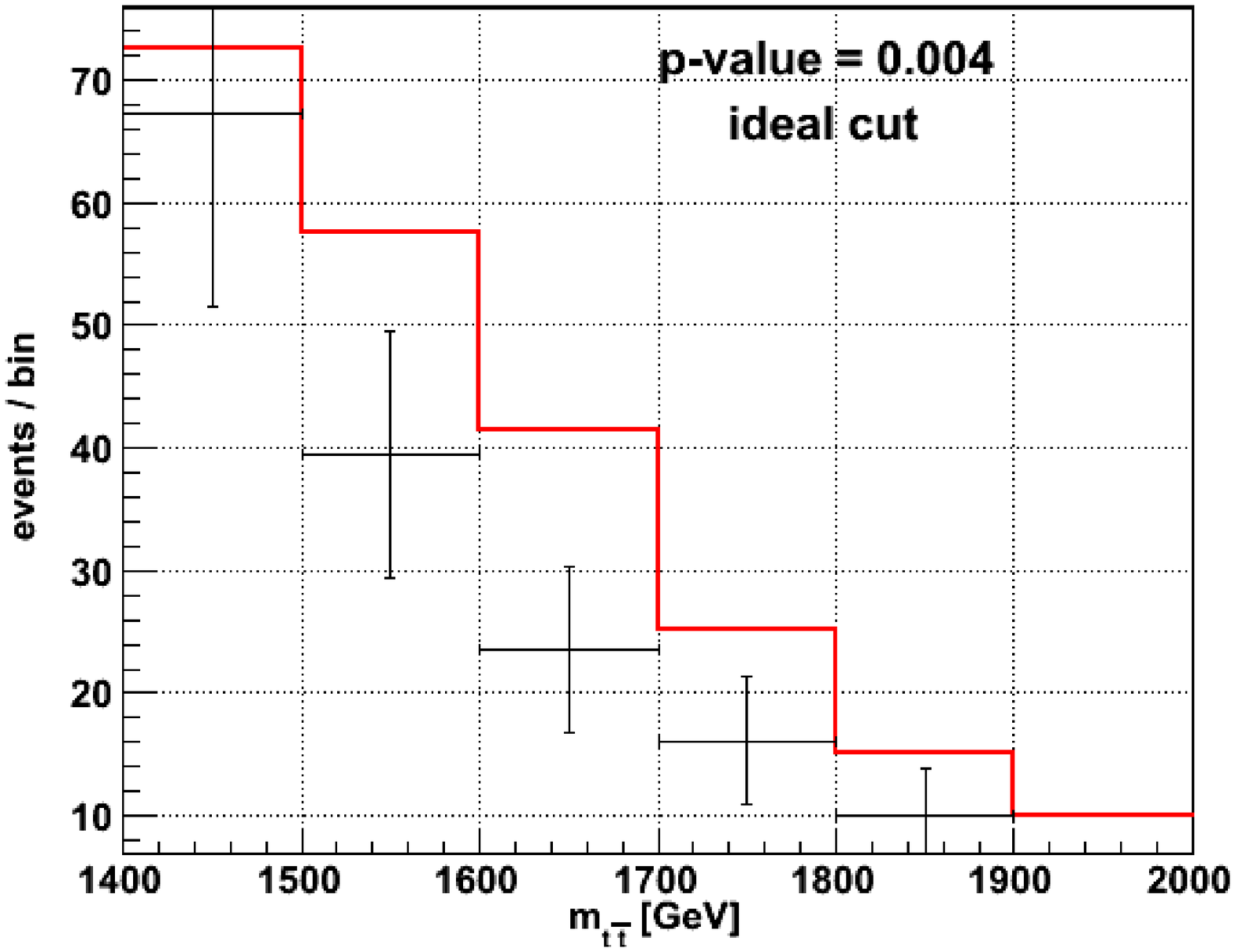}
\newline
(d)
\end{center}
\end{minipage}
\caption{[color online] \ttbar invariant mass spectrum for the $30$ fb$^{-1}$ expected LHC sample.  The (black) histogram with error bars corresponds to only SM, and the (red) histogram corresponds to SM+NP.  In panel (a) we plot both spectrum without applying any cut on the sample, in (b) we apply the best cut that can be achieved in using $\beta$ and $p_T$, in (c) the same as in (b) but using $\beta$, $p_T$ and $\theta$, and finally (d) corresponds to an ideal cut, as explained in the text.  In each panel we determine the significance of the deviation according to a $\chi^2$ test (see text). }
\label{histograms}
\end{figure}

We take as an example a benchmark model with a gluon resonance of $1.5$ TeV reminiscent of strongly coupled/composite models. A characteristic feature of this kind of models is that this massive gluon couples stronger to the heavy than to the light quarks, and it cannot be created from a fusion of two QCD gluons. Thus, this kind of models can illustrate and test the studied cuts. The interaction with SM particles is:
\bea
\mathcal L_{G'} =  g_s \sum_{q_i}  f_{q} \, \bar q \Gslash' q 
\eea 
\noindent where $q$ stands for all quarks. We take $f_{(u,d,s,c)_{L,R}} = 0.025$, $f_{t_R} = 4.6$ and all other couplings are set to zero.  With this choice of parameters the width of $G'$ yields $\Gamma_{G'} = 290$ GeV, and the resonance would not be visible with the LHC running at $8$ TeV and an accumulated luminosity of $30$ fb$^{-1}$ without any cut to improve the \fqq fraction.   

With the same computational tools used in the previous section, we have simulated a $30$ fb$^{-1}$ sample of $pp\to t\bar t$ and assumed a reference selection efficiency of $6\ \%$ \cite{ac2}. We have simulated only SM and jointly SM+NP, to take into account the interference coming from the large width effects. 
In Fig.~\ref{histograms} we show the comparison in the region of interest of the expected $m_{t\bar t}$-spectrum of only SM (histogram with error bars, in black) and SM+NP (histogram, in red) for: ($a$) no cuts at all; ($b$) cuts in $\beta$ and $p_T$; ($c$) cuts in $\beta$, $p_T$ and $\theta$; and $(d)$ an ideal cut where only the quark-annihilation events in the sample are left.  The error bars correspond to statistical and systematic uncertainties, as discussed in the previous section.   We choose the cuts in plots $(b)$ and $(c)$ from a scanning that looks for the cuts that maximize the significance of a $\chi^2$ test to 3 bins of $100$ GeV width from $1500$ GeV to $1800$ GeV.  We have checked that if we had chosen the {\it a priori} cuts from the previous section SM analysis to the $1600-1800$ GeV bin, then we would have obtained an analogous, but not optimal, improvement.

\begin{table}[!htb]
\begin{center}
\begin{tabular}{|c|c|l|l|l|l|}
\hline
plot&cuts & $f_s$ & \fqq & p-value  & $\sigma$ \\
\hline
$(a)$ & w/o cuts & 1.00 & 0.41 & 0.36 & 0.4 \\ 
$(b)$ & $\beta > 0.45$, $p_T < 75$ GeV & 0.20 & 0.63 & 0.13 & 1.1  \\
$(c)$ & $\beta > 0.08$,  $p_T < 65$, $\lvert \cos \theta \rvert < 0.66$ GeV & 0.22 & 0.75 & 0.02 & 2.1  \\
$(d)$ & ideal cut & 0.41 & 1.00 & 0.004 & 2.7 \\
\hline
\end{tabular} 
\caption{\label{table}[color online] Set of cuts used to differentiate SM+NP from SM using different set of variables.  The first two columns indicate the plot in Fig.~\ref{histograms} and the performed cuts, the third and fourth columns show their effect on the sample ($f_s$ and $f_{q\bar q}$), and the following columns the expected differentiation in computing a $\chi^2$ test, as explained in the text.}
\end{center}
\end{table}

To quantify the gradual improvement in resolving the resonance in Fig.~\ref{histograms}, we show in Table \ref{table} the $f_s$, \fqq, $p$-values for the $\chi^2$ tests, and the Gaussian-equivalent significance ($\sigma$) of all the set of cuts in the figure.  We also show the value of the cuts for Fig.~\ref{histograms}b and Fig.~\ref{histograms}c.  As we can determine from the table (second and third row), a cut in $\beta$ is not too tight at these energies and luminosities.  Moreover, in passing from second to third row, when the cut in $\theta$ is included, we see that the cut in $\beta$ is practically replaced by a cut in $\theta$.  We see in this example that this exchange of cuts increases $f_s$ by $\sim 10\ \%$ and \fqq by $\sim 20\ \%$, which yields an improvement in the $p$-value from $0.13$ to $0.02$. 

We have repeated the same computations as in this section for other models of resonant NP and obtained similar results.  For the best cuts in the analogous to Fig.~\ref{histograms}c we always obtain $\beta \gtrsim 0-0.1$, $|\cos(\theta)| \lesssim 0.65-0.85$ and $p_T \lesssim 30-70$ GeV.  This is in agreement with the previous section analysis.  It could be valuable to stress at this point that the results in this section have the only purpose of showing the qualitative effect of these cuts in the increment in the sensitivity through a specific example.  

\section{Discussions}

The aim of this section is to relate the present analysis with previous works. Besides $p_T$ and $\beta$, the rapidity of the top and spin correlations of its decays also have been used to separate NP and SM $q \bar q \to t \bar t$ production from the $gg \to t \bar t$ background. First, we discuss in this section some differences among the present analysis that makes focus on $\theta$ as a suitable kinematic variable and previous studies where the rapidity $y$ is used to impose tight selection cuts. And finally, we discuss briefly about the different regimes where $\theta$ cuts and spin correlation are useful to improve the \fqq fraction. 

It is well known that there is a correlation between the center of mass angle $\theta$ and the difference of rapidity $\Delta y = y^t - y^{\bar t}$,
\bea
\Delta y  = \log \frac{1+\beta_t \cos \theta}{1-\beta_t \cos \theta} \, ,
\eea
\noindent where $\beta_t = \sqrt{ 1-4 m_t^2 /\hat s }$. The partonic energy $\hat s$ entering this relation may dilute the correlation between this two variables. Only if $\beta_t$ is fixed, there is a one-to-one correspondence between $\theta$ and $\Delta y$. However, if we look at a bin of invariant mass where $\beta_t$ has large variations then there could be a lower correlation between them. For instance, if we consider the \mtt bin of 1600 to 1800 GeV, $\Delta y$ varies up to $4\ \%$ when $\theta$ is fixed. Therefore, for a high \mtt bin, cuts on $\theta$ and $\Delta y$ are roughly equivalent. Moreover, if we take several bins in the high \mtt region, for the same cut on $\theta$, we will have a different (but equivalent) cut on $\Delta y$ in each bin. On the other hand, close to the threshold, $\beta_t$ varies greatly and the correlation is more diluted. Thus, within the 400-425 GeV bin, $\Delta y$ varies up to $20\ \%$ for fixed values of $\theta$. 

Cuts on $|\Delta y|$ have already been considered in the literature \cite{bai,as,deltay}. However, they have been applied at low energies and for certain class of models. For instance, in $Z'$ flavor violating models, \ttbar are produced through the exchange of the $Z'$ in the $t$-channel and then it produces \ttbar events mostly in the forward region. This situation, which does not correspond to resonant NP, is quite the opposite to the present analysis. For such models, it has been shown that a lower cut on $|\Delta y|$ enhances the sensitivity to the new interactions. Since these interactions arise at low $m_{t\bar t}$, there is not a considerable increment of the gluon-fusion fraction that may spoil the enhancement generated by the new interactions. In cases where the new physics contributions are significant at higher $m_{t\bar t}$, we have to deal with an interplay between increasing the NP $t$-channel contribution and decreasing the sensitivity with the increment of the gluon-fusion \ttbar production.

The rapidity has also been used in the context of the charge asymmetry in \ttbar production at the LHC \cite{hewett,gr}. In those cases, the basic idea is also to discriminate \qqbar from $gg$-fusion events by imposing different cuts on $y^t$ and $y^{\bar t}$. The procedure traces the discrimination back to the correlation between $y$ and $\beta$. However, the angular variable $\theta$ (or $\Delta y$) plays no role in that method.
	
Finally, for many of the models that have been proposed to accommodate the $A_{FB}$, there exist a high \mtt resonance that induces an interference at low $m_{t\bar t}$, for instance, a heavy axigluon that has been integrated out \cite{maltoni2}. Therefore, the new physics contribution arise at low \mtt and this may seem to be the reason why $\theta$ has never been taken as a relevant kinematic variable in previous studies.   

Spin correlations in the production and decay of tops have been extensively discussed \cite{parke,bai,spin}. The aim of those studies is to compare predictions made within the SM or its extensions with the measured angular distributions of the decay products of $t$ and $\bar t$. In general, different models give place to distinct angular distribution patterns and the measurement of spin correlations may provide additional and relevant information on the structure of possible new interactions. Such spin correlations are optimized when the asymmetry between like- and unlike-helicities reaches certain maximum value. Usually, for a given axis from which the spin projection is defined, one seeks to perform kinematic cuts that increase this asymmetry. This in turn allows to discriminate like- and unlike-helicities. 

From Eqs.~(\ref{qqLL})-(\ref{ggLR}) and their corresponding discussion in Section 2 we have learned that angular distribution for both quark-annihilation and $gg$-fusion production mechanisms are essentially flat close to the threshold. Therefore, $\theta$ cuts cannot be used to improve the \qqbar fraction. On the contrary, a distinction based on spin correlation is envisaged since only like-helicity $gg \to t \bar t$ production is possible whereas the two spin production mechanism are present for $q \bar q \to t \bar t$ as we can see from the Eqs.~(\ref{qqLL})-(\ref{ggLR}). The opposite situation we encounter at large \mtt. There, only unlike-helicity $q \bar q \to t \bar t$ and $gg \to t \bar t$ production mechanisms are possible, being hardly expected that spin correlation being useful to distinguish \qqbar from $gg$-fusion production contrary to $\theta$ cuts that has been proven to be suitable to increase the \qqbar fraction.    

\section{Conclusions}

We have studied the collider phenomenology of $t\bar t$ resonance
searches above 1 TeV at 30 fb$^{-1}$.  We have focused our analysis 
on resonances that couple exclusively to quarks
which, at the LHC, would be naturally hidden behind the large
gluon-fusion $t \bar t$ production.

The usual variables to enlarge the quark-annihilation fraction \fqq
of a $t\bar t$ sample are the longitudinal and transverse momentum of
the $t\bar t$ pair, $\beta$ and $p_T$, respectively. In addition to these variables, 
we exploit the fact that for masses above 1 TeV the large boost of the tops 
in the center of mass system yields a correlation between their center of mass
scattered angle $\theta$ and the initial partons of the event (this
correlation is suppressed for invariant masses below 1 TeV).  Since the origin
of the relationship between the initial state partons and $\beta$ is
the proton PDF, whereas for $p_T$ is the initial state
radiation, and for $\theta$ the dynamics of the partonic collision,
there are not a priori reasons to expect these variables to be considerably
correlated, and all three should be used together to texture the
initial state partons of an event.

We have simulated $t\bar t$ production in the large invariant mass
regime and scanned over different set of cuts on the variables mentioned above
to increase the quark-annihilation fraction of the sample.
Given a set of invariant mass bins where the NP resonance could be
expected, the optimal cut is the one that enhances the most the
significance of a resonance search over the SM background. This is
achieved when a set of cuts increases \fqq while keeping the selected fraction $f_s$ still large since, 
for resonances beyond 1 TeV, a critical reduction in the amount of 
events could spoil the signal due to an increase of the statistical uncertainties.

Our results show that $\beta$ is not as useful as
expected in this resonance search.  The reason for this is that this
variable requires tight cuts in order to yield significant increases
in $f_{q\bar q}$, and there are not too many events in the large invariant
mass bins in the 2012 data.  On the other hand, $\theta$ does not
reach large values of $f_{q\bar q}$, but it has an important increase as one
begins cutting the sample.  Hence, the reasons why $\theta$ is useful
in the large invariant mass bins are two-folded: the variable requires
relativistic tops to distinguish the initial state partons, and in
this region the cuts should be moderate.  At last, the cut in
$p_T$ is always useful for any invariant mass bin.

We have studied a reference case of a $1.5$ TeV resonance and found that the
optimal cuts in these three parton level variables is in agreement with
the previous sections analyses.  We have also shown how the
significance of the resonance search behaves as one gradually includes
all variables in the selection cuts.  Summarizing, when we include all three variables, we found that
$\beta$ is not relevant, and $p_T$ and $\theta$ provide the most
useful cuts to enhance the significance of the resonance search.

The results in this article are a guideline to point the searches of NP 
resonances above 1 TeV for 30 fb$^{-1}$ at a center of mass energy of $8$ TeV. It is important to stress the
perspectives for upcoming luminosity and energy increases. For NP resonances coupled to quarks
and masses above 1 TeV, a rise in the luminosity will make $\theta$ 
remain as a useful variable, whereas $\beta$ will improve as a \qqbar discriminator too.  On the other hand, a rise of the center of mass energy, will allow the exploration of heavier resonances within this framework.

\section*{Acknowledgments}

We thank L.~Da Rold for useful conversations.  This work has been partially supported by ANPCyT (Argentina) under
grants PICT-PRH 2009-0054 (A.S) and PICT 2011-0359 (A.E. and J.I.S.V.)  and by CONICET (Argentina).

{}


\begin{thebibliography}{}

\bibitem{higgs} 
G.~Aad {\it et al.}  [ATLAS Collaboration],
  Phys.\ Lett.\ B {\bf 716}, 1 (2012)
  [arXiv:1207.7214 [hep-ex]];
 S.~Chatrchyan {\it et al.}  [CMS Collaboration],
  Phys.\ Lett.\ B {\bf 716}, 30 (2012)
  [arXiv:1207.7235 [hep-ex]].

\bibitem{susy}
https://twiki.cern.ch/twiki/bin/view/AtlasPublic/SupersymmetryPublicResults;
\newline
https://twiki.cern.ch/twiki/bin/view/CMSPublic/PhysicsResultsSUS .

\bibitem{tev}
 The CDF Collaboration, Conf.~Note 10807, http://www-cdf.fnal.gov/physics/new/top/ 2012/LepJet\_AFB\_Winter2012/CDF10807.pdf;
 A.~Lister [CDF and D0 Collaboration],
  arXiv:0810.3350 [hep-ex];
 T.~Aaltonen {\it et al.}  [CDF Collaboration],
  Phys.\ Rev.\ Lett.\  {\bf 101}, 202001 (2008)
  [arXiv:0806.2472 [hep-ex]];
V.~M.~Abazov {\it et al.}  [D0 Collaboration],
  Phys.\ Rev.\ D {\bf 84}, 112005 (2011)
  [arXiv:1107.4995 [hep-ex]];
  T.~Aaltonen {\it et al.}  [CDF Collaboration],
  Phys.\ Rev.\ D {\bf 83}, 112003 (2011)
  [arXiv:1101.0034 [hep-ex]].
  
\bibitem{ac}
  arXiv:1203.4211 [hep-ex];
S.~Chatrchyan {\it et al.}  [CMS Collaboration],
  Phys.\ Lett.\ B {\bf 709}, 28 (2012)
  [arXiv:1112.5100 [hep-ex]];
    G.~Aad {\it et al.}  [ATLAS Collaboration],
  Eur.\ Phys.\ J.\ C {\bf 72}, 2039 (2012)
  [arXiv:1203.4211 [hep-ex]].

\bibitem{ac2}
  S.~Chatrchyan {\it et al.}  [CMS Collaboration],
  Phys.\ Lett.\ B {\bf 717}, 129 (2012)
  [arXiv:1207.0065 [hep-ex]].
  
\bibitem{few}
 R.~Barcelo, A.~Carmona, M.~Masip and J.~Santiago,
  Phys.\ Lett.\ B {\bf 707}, 88 (2012)
  [arXiv:1106.4054 [hep-ph]];
 E.~Alvarez, L.~Da Rold, J.~I.~S.~Vietto and A.~Szynkman,
  JHEP {\bf 1109}, 007 (2011)
  [arXiv:1107.1473 [hep-ph]];
    G.~Marques Tavares and M.~Schmaltz,
  Phys.\ Rev.\ D {\bf 84}, 054008 (2011)
  [arXiv:1107.0978 [hep-ph]];
    E.~Alvarez and E.~C.~Leskow,
  arXiv:1209.4354 [hep-ph];
  J.~Drobnak, A.~L.~Kagan, J.~F.~Kamenik, G.~Perez and J.~Zupan,
  arXiv:1209.4872 [hep-ph].
  
\bibitem{acsearches}
  Y.~-k.~Wang, B.~Xiao and S.~-h.~Zhu,
  Phys.\ Rev.\ D {\bf 82}, 094011 (2010)
  [arXiv:1008.2685 [hep-ph]];
  Y.~-k.~Wang, B.~Xiao and S.~-h.~Zhu,
  Phys.\ Rev.\ D {\bf 83}, 015002 (2011)
  [arXiv:1011.1428 [hep-ph]];
  B.~Xiao, Y.~-K.~Wang, Z.~-Q.~Zhou and S.~-h.~Zhu,
  Phys.\ Rev.\ D {\bf 83}, 057503 (2011)
  [arXiv:1101.2507 [hep-ph]];
  J.~-F.~Arguin, M.~Freytsis and Z.~Ligeti,
  Phys.\ Rev.\ D {\bf 84}, 071504 (2011)
  [arXiv:1107.4090 [hep-ph]].
  
\bibitem{hewett}
J.~L.~Hewett, J.~Shelton, M.~Spannowsky, T.~M.~P.~Tait and M.~Takeuchi,
  Phys.\ Rev.\ D {\bf 84}, 054005 (2011)
  [arXiv:1103.4618 [hep-ph]].

\bibitem{bai}
  Y.~Bai and Z.~Han,
  JHEP {\bf 1202}, 135 (2012)
  [arXiv:1106.5071 [hep-ph]].
  
\bibitem{as}
  J.~A.~Aguilar-Saavedra, A.~Juste and F.~Rubbo,
  Phys.\ Lett.\ B {\bf 707}, 92 (2012)
  [arXiv:1109.3710 [hep-ph]].

\bibitem{gr}
  J.~H.~Kuhn and G.~Rodrigo,
  JHEP {\bf 1201}, 063 (2012)
  [arXiv:1109.6830 [hep-ph]].

\bibitem{afbseq} 
  E.~Alvarez,
  Phys.\ Rev.\ D {\bf 85}, 094026 (2012)
  [arXiv:1202.6622 [hep-ph]].

\bibitem{mttagashe}
 K.~Agashe, A.~Belyaev, T.~Krupovnickas, G.~Perez and J.~Virzi,
  Phys.\ Rev.\ D {\bf 77}, 015003 (2008)
  [hep-ph/0612015].

\bibitem{mttseq}
  E.~Alvarez,
  Phys.\ Rev.\ D {\bf 86}, 037501 (2012)
  [arXiv:1205.5267 [hep-ph]].
  
\bibitem{barger}
  V.~Barger, T.~Han and D.~G.~E.~Walker,
  Phys.\ Rev.\ Lett.\  {\bf 100}, 031801 (2008)
  [hep-ph/0612016].

\bibitem{maltoni}
  R.~Frederix and F.~Maltoni,
  JHEP {\bf 0901}, 047 (2009)
  [arXiv:0712.2355 [hep-ph]].

\bibitem{dijet}
  S.~Chatrchyan {\it et al.}  [CMS Collaboration],
  Phys.\ Lett.\ B {\bf 704}, 123 (2011)
  [arXiv:1107.4771 [hep-ex]];
   [ATLAS Collaboration],
  ATLAS-CONF-2012-038.

\bibitem{parke}
  G.~Mahlon and S.~J.~Parke,
  Phys.\ Rev.\ D {\bf 53}, 4886 (1996)
  [hep-ph/9512264].

\bibitem{spin2}
 B.~Grinstein, C.~W.~Murphy, D.~Pirtskhalava and P.~Uttayarat,
  JHEP {\bf 1208}, 073 (2012)
  [arXiv:1203.2183 [hep-ph]].

\bibitem{mgme}
 J.~Alwall, P.~Demin, S.~de Visscher, R.~Frederix, M.~Herquet, F.~Maltoni, T.~Plehn, D.~L.~Rainwater {\it et al.},
  JHEP {\bf 0709 } (2007)  028,
  [arXiv:0706.2334 [hep-ph]].

\bibitem{pythia}
  T.~Sjostrand, S.~Mrenna and P.~Z.~Skands,
  JHEP {\bf 0605} (2006) 026
  [arXiv:hep-ph/0603175].
  
\bibitem{mlm}
  J.~Alwall, S.~Hoche, F.~Krauss, N.~Lavesson, L.~Lonnblad, F.~Maltoni, M.~L.~Mangano and M.~Moretti {\it et al.},
  Eur.\ Phys.\ J.\ C {\bf 53}, 473 (2008)
  [arXiv:0706.2569 [hep-ph]].

\bibitem{mcfm}
  [CMS Collaboration],
  CMS-PAS-TOP-12-007.
  J.~M.~Campbell and R.~K.~Ellis,
  Nucl.\ Phys.\ Proc.\ Suppl.\  {\bf 205-206}, 10 (2010)
  [arXiv:1007.3492 [hep-ph]].
       
\bibitem{20p}
  [CMS Collaboration],
  CMS-PAS-TOP-11-009.
  
\bibitem{js}
  M.~Backovic and J.~Juknevich,
  arXiv:1212.2978 [hep-ph];
 J.~Gallicchio and M.~D.~Schwartz,
  arXiv:1211.7038 [hep-ph],
  and reference therein.
  
\bibitem{deltay}    
 S.~Berge and S.~Westhoff,
  arXiv:1208.4104 [hep-ph].
  
\bibitem{maltoni2}
  C.~Degrande, J.~-M.~Gerard, C.~Grojean, F.~Maltoni and G.~Servant,
  JHEP {\bf 1103}, 125 (2011)
  [arXiv:1010.6304 [hep-ph]].

\bibitem{spin}
   G.~Mahlon and S.~J.~Parke,
  Phys.\ Lett.\ B {\bf 411}, 173 (1997)
  [hep-ph/9706304];
    T.~Stelzer and S.~Willenbrock,
  Phys.\ Lett.\ B {\bf 374}, 169 (1996)
  [hep-ph/9512292];
and many others.

\end{thebibliography}
\end{document}